\documentclass{article}
\usepackage{arxiv}
\usepackage[utf8]{inputenc} % allow utf-8 input
\usepackage[T1]{fontenc}    % use 8-bit T1 fonts
\usepackage{url}            % simple URL typesetting
\usepackage{booktabs}       % professional-quality tables
\usepackage{amsfonts}       % blackboard math symbols
\usepackage{amsmath}
\usepackage{amssymb}
\usepackage{siunitx}
\usepackage{nicefrac}       % compact symbols for 1/2, etc.
\usepackage{microtype}      % microtypography
\usepackage{lipsum}		% Can be removed after putting your text content
\usepackage{doi}
\usepackage[numbers]{natbib}
\usepackage[normalem]{ulem}
\usepackage{graphicx}
\usepackage{mathtools}
\usepackage{xcolor} 
\usepackage{wasysym}
\usepackage{cuted}
\usepackage{hyphenat}

\usepackage{bookmark}

\makeatletter
\def\ps@myheadings{%
    \let\@oddfoot\@empty\let\@evenfoot\@empty
    \def\@evenhead{\thepage\hfil\slshape\leftmark}%
    \def\@oddhead{{\slshape\rightmark}\hfil\thepage}%
    \let\@mkboth\@gobbletwo
    \let\sectionmark\@gobble
    \let\subsectionmark\@gobble
    }
  \if@titlepage
  \renewcommand\maketitle{\begin{titlepage}%
  \let\footnotesize\small
  \let\footnoterule\relax
  \let \footnote \thanks
  \null\vfil
  \vskip 60\p@
  \begin{center}%
    {\LARGE \@title \par}%
    \vskip 3em%
    {\large
     \lineskip .75em%
      \begin{tabular}[t]{c}%
        \@author
      \end{tabular}\par}%
      \vskip 1.5em%
    {\large \@date \par}%       % Set date in \large size.
  \end{center}\par
  \@thanks
  \vfil\null
  \end{titlepage}%
  \setcounter{footnote}{0}%
}
\else
\renewcommand\maketitle{\par
  \begingroup
    \renewcommand\thefootnote{\@fnsymbol\c@footnote}%
    \def\@makefnmark{\rlap{\@textsuperscript{\normalfont\@thefnmark}}}%
    \long\def\@makefntext##1{\parindent 1em\noindent
            \hb@xt@1.8em{%
                \hss\@textsuperscript{\normalfont\@thefnmark}}##1}%
    \if@twocolumn
      \ifnum \col@number=\@ne
        \@maketitle
      \else
        \twocolumn[\@maketitle]%
      \fi
    \else
      \newpage
      \global\@topnum\z@   % Prevents figures from going at top of page.
      \@maketitle
    \fi
    \thispagestyle{plain}\@thanks
  \endgroup
  \setcounter{footnote}{0}%
}
\makeatother

\newcommand{\Figref}[1]{Fig.~\ref{#1}}
\newcommand{\Figrefa}[1]{Fig.~\ref{#1}(a)}
\newcommand{\Figrefb}[1]{Fig.~\ref{#1}(b)}
\newcommand{\Figrefc}[1]{Fig.~\ref{#1}(c)}
\newcommand{\Figrefd}[1]{Fig.~\ref{#1}(d)}
\newcommand{\Figrefe}[1]{Fig.~\ref{#1}(e)}
\newcommand{\Figreff}[1]{Fig.~\ref{#1}(f)}

\newcommand{\onlinecite}[1]{[\hspace{-0.7 ex} \nocite{#1}\citenum{#1}\hspace{0.2 ex}]} % from https://latex.org/forum/viewtopic.php?t=8897
\DeclareMathOperator{\sgn}{sgn}

\title{Terahertz detection with graphene FETs: photothermoelectric and resistive self-mixing contributions to the detector response}

\renewcommand{\shorttitle}{Terahertz detection with graphene FETs: \\ photothermoelectric and resistive self-mixing contributions to the detector response}

\author{Florian Ludwig\\
Physikalisches Institut\\
Johann Wolfgang Goethe-Universität\\
DE-60438 Frankfurt am Main, Germany \\
\texttt{ludwig@physik.uni-frankfurt.de} \\
   \And
Andrey Generalov \\
VTT Technical Research Centre of Finland\\
FI-02044 Espoo, Finland\\
\texttt{andrey.generalov@vtt.fi} \\
  \And
Jakob Holstein \\
Physikalisches Institut\\
DE-60438 Frankfurt am Main, Germany\\
   \And
Anton Murros \\
VTT Technical Research Centre of Finland\\
FI-02044 Espoo, Finland \\
   \And
Klaara Viisanen \\
VTT Technical Research Centre of Finland\\
FI-02044 Espoo, Finland \\
   \And
Mika Prunnila \\
VTT Technical Research Centre of Finland\\
FI-02044 Espoo, Finland \\
   \And
Hartmut G. Roskos \\
Physikalisches Institut\\
DE-60438 Frankfurt am Main, Germany\\
\texttt{roskos@physik.uni-frankfurt.de} \\
}

\hypersetup{
pdftitle={Terahertz detection with graphene FETs: photothermoelectric and resistive self-mixing contributions to the detector response},
pdfsubject={physics.app-ph, physics.cond-mat},
pdfauthor={Florian Ludwig, Andrey Generalov, Jakob Holstein, Anton Murros, Klaara Viisanen, Mika Prunnila, Hartmut G. Roskos},
pdfkeywords={graphene, field-effect transistor, terahertz detection, TeraFET, photodetector, photothermoelectric effect, resistive selfmixing},
}

\begin{document}
\begin{titlepage}
\maketitle
\vspace{-1em}
\begin{abstract}
\vspace{-1.0em}
% ***** New *******
\textcolor{black}{Field-effect transistors coupled to integrated antennas (TeraFETs) are photodetectors being actively developed for the THz frequency range ($\sim$100~GHz – 10 THz). Among them, Graphene TeraFETs (G-TeraFETs) have demonstrated distinctive photoresponse features compared to those made from elementary semiconductors. For instance, previous studies have shown that G-TeraFETs exhibit a THz response that comprises two components: the resistive self-mixing (RSM) and photothermoelectric effect (PTE). The RSM and PTE arise from carrier density oscillations and carrier heating, respectively. In this work, we confirm that the photoresponse can be considered a combination of RSM and PTE, with PTE being the dominant rectification mechanism at higher frequencies. For our CVD G-TeraFETs with asymmetric antenna coupling, the PTE response dominates over the RSM at frequencies above 100 GHz. We find that relative contribution of RSM and PTE to the photoresponse is strongly frequency dependent. Electromagnetic wave simulations show that this behavior is due to the relative change in the total dissipated power between the gated and ungated channel regions of the G-TeraFET as the frequency increases. The simulations also indicate that the channel length over which the PTE contributes to the photoresponse below the gate electrode is approximately the same as the electronic cooling length. Finally, we identify a PTE contribution that can be attributed to the contact doping effect in graphene close to the metal contacts. Our detectors achieve a minimum optical noise-equivalent power of 101 (114) pW/$\sqrt{\rm{Hz}}$ for asymmetric (symmetric) THz antenna coupling conditions at 400 GHz. This work demonstrates how the PTE response can be used to optimize the THz responsivity of G-TeraFETs.}
% ***** OLD *******
%THz field-effect transistor (TeraFET) devices are antenna-coupled photodetectors that are being actively developed for the THz frequency range ($\sim$ 100~GHz – 10 THz). Among them, Graphene TeraFETs (G-TeraFETs) have shown some unique features in their photoresponse compared to those made from elementary semiconductors. For example, it has been shown that the THz response of G-TeraFETs consists of two components: the so-called resistive self-mixing (RSM) and photothermoelectric effect (PTE), which result from carrier density oscillations and carrier heating, respectively. Here we explicitly demonstrate that the response can be treated as a combination of RSM and PTE and that the PTE is the dominant recitication mechanism at higher THz frequencies. In the case of our CVD G-TeraFETs asymmetrically coupled to the THz antenna, the PTE response starts to dominate over the RSM already above 100 GHz photon field frequencies. By electromagnetic wave simulations, we show that this behavior is a consequence of the relative change in the total dissipated THz power between the gated and ungated regions of the G-TeraFET. We also identify a PTE contribution which can be attributed to the Fermi level gradient due to the graphene – metal contacts. Our detectors achieve a minimum optical noise-equivalent power of 101 (114) pW/$\sqrt{\rm{Hz}}$ for asymmetric (symmetric) THz antenna coupling conditions at 400 GHz. Our work shows how the PTE response can be used to optimise the THz responsivity of G-TeraFETs.
\end{abstract}
\vspace{-1.2em}

\end{titlepage}

\flushbottom

\pagebreak 

\section{Introduction}

Device physics and technologies in the THz frequency range (roughly defined as 100~GHz to 10~THz) have attracted a considerable research interest over the last decades. The technological exploitation of this frequency range is primarily driven by a variety of applications\cite{Roadmap2021}, such as security screening \cite{Appleby2007,Shen2005,Friederich}, quality control and non-destructive testing \cite{Ahi2018,Nüßler2021,MarioWatsonPreu2018,Hasegawa2003}, spectroscopy of molecular compounds in atmosphere and space \cite{Veeraselvam2021,Neese2012,Hsieh2016}, biomedical sensing and imaging \cite{Bowman2016,Taylor2011}, and high-speed communications \cite{Harter2020,Shur2021}. The THz range is also fertile ground for fundamental studies of carrier dynamics \cite{Meng2015,Ulbricht2011} and quasi-particle physics in general  \cite{Pashnev2020,Thomson2017,Kaindl2009}. Access to the THz frequency range and applications requires tunable and powerful sources and fast and sensitive detectors. These should be cost effective, compact, integrable into arrays, and compatible with room temperature operation and mainstream semiconductor manufacturing processes. In this context, THz field-effect transistors (TeraFETs) have evolved from the early proofs of concept \cite{Knap2009, Vicarelli2012, Koppens2014} to a promising family of THz detectors, competing with the more established diode-based detectors \cite{Mehdi2017,Ahmad2019,Yadav2023}. In particular, they achieve high sensitivities with optical noise-equivalent powers (NEPs) as low as 20~pW/$\surd$Hz \cite{Mateos2020} (see also table~\ref{tab:NEPvalues}) and sub-nanosecond detection speed \cite{Viti2021,Ikamas-pico,Preu2012}, which also makes them suitable for coherent heterodyne detection \cite{Wiecha2021,GrzybTHzDirectDetector,LisauskasSubharmonic}. 

Over the last fifteen years, TeraFETs based have been developed and studied, initially focusing on classical semiconductor material systems such as Si \cite{Hillger2020,Ikamas2018,Lisauskas2009}, AlGaN/GaN\cite{Bauer2019,Sun2020,Boppel2016,Sun2012}, AlGaAs/GaAs \cite{Regensburger2018_CW,Nagatsuma2013,Dyer2012}, and InGaAs/GaAs  \cite{Popov2011,Knap2009,Fat2006}. Si MOSFET based devices continue to be among the TeraFET detectors with the highest sensitivity at room temperature over most of the THz spectral range\cite{Javadi2021,Ferreras2021,Bauer2019,Ikam2017}. They benefit from mature CMOS fabrication, which ensures high fabrication yields and excellent reproducibility of the detector performance, conditions that make them well suited for implementation in cameras and integrated systems\cite{Hillger2020,Zda15,Had12}. More recently, TeraFET studies have been extended to 2D van der Waals materials such as mono- and bilayer graphene\cite{Delgado-Notario2022,Viti2021,Viti2020,Bandurin2018a,Bandurin2018b,Generalov2017,Zak2014,Cai2014,Vicarelli2012}, 
black phosphorus\cite{Viti2015}, and Bi$_2$Te$_{3-x}$Se$_x$\cite{Viti2016}, with the hope of achieving improved detector performance by exploiting the unique properties of 2D materials, such as high carrier mobility. For G-TeraFETs, significant efforts have already been made to improve the detector sensitivity, e.g. by employing exfoliated monocrystalline graphene and encapsulating it in layers of monocrystalline hexagonal boron nitride (hBN) to reduce interface and phonon scattering. The best reported NEP value for a THz detector based on van der Waals materials to date is for a thermoelectric graphene detector. A NEP value of 80~pW/$\surd$Hz (\textit{cross-sectional} NEP \footnote{One distinguishes between \textit{optical} and \textit{cross-sectional} responsivity and NEP\cite{Bauer2019,Bauer2016}. The optical voltage or current responsivity is obtained as the measured rectified voltage or current divided by the (total) power of the THz beam measured in front of the detector. The cross-sectional current or voltage responsivity is used when the antenna cross-section is smaller than the beam size. The cross-sectional responsivity is calculated as the optical responsivity multiplied by the ratio of the (measured or calculated) beam cross-sectional area at the antenna to the (calculated) antenna cross-sectional area\cite{Javadi2021}.}) has been obtained at 2.5~THz for operation at room temperature (see table~\ref{tab:NEPvalues}).

The underlying THz detection mechanism of TeraFETs made from classical semiconductor materials is typically resistive self-mixing (RSM), which is induced in the two-dimensional electron gas (2DEG) of the FET channel \cite{Lisauskas2013,Boppel2012,Ojefors2009,Lisauskas2009} by the THz radiation asymmetrically injected into the channel from the source (S) or drain (D) contact. This results in a rectified drain-to-source current (or voltage) proportional to the power of the THz signal. At low frequencies, the mixing is described by a quasi-static response of the FET, but at high millimeter-wave frequencies and for THz waves the channel can no longer be treated as a lumped element but must be considered as a non-linear waveguide. The rectification process is then often referred to as \textit{distributed resistive self-mixing}\cite{Boppel2012} and the efficiency of the rectification by RSM is expected to increase in the THz frequency range due to the development of plasma waves (plasmons) in the gated region of the 2DEG. 

Interestingly, already during the first realization and characterization of G-TeraFETs, it became clear that, besides RSM, the so-called photothermoelectric effect (PTE) contributes significantly as a second rectification mechanism \cite{Zak2014,Vicarelli2012,Bauer2015}. The PTE results from local heating of the carriers due to the THz field. The carriers remain thermally decoupled from the crystal lattice over relatively long time scales, as the electron-phonon relaxation pathway \textcolor{black}{of the carrier ensemble, which is only weakly heated by the THz radiation, is reduced to predominantly slow processes such as} acoustic phonon scattering ($\sim$1-2~ps)\cite{Poba2021,Song2015,Brida2013,Tomadin2013,Song2011} (\textcolor{black}{the emission of optical phonons being strongly limited by their high energy of 160~meV\cite{Pop2012}, to be compared with the thermal energy scale of $k_B T \approx 25$~meV}). Carrier-carrier scattering takes place on time scales of $\le$20~fs\cite{Tomadin2018,Meng2016,Gierz2013,Johannsen2013} leading to a fast re-establishment (on sub-100-fs time scales) of a thermal carrier distribution at temperature $T_C > T_L$ ($T_L$ is the lattice temperature). When carriers in graphene are photo-excited inhomogeneously, the resulting spatial gradients in carrier temperature lead to thermoelectric voltages. In contrast to the classical Seebeck effect, for which $T_C \approx T_L$, this is a purely electronic thermoelectric effect, which has led to the term \textit{hot-carrier thermoelectric effect}, or PTE in the case of photon excitation drive \cite{Cai2014, Jung2016, Bandurin2018b, Castilla2019}.

Since G-TeraFETs can in principle benefit from two rectification mechanisms being active simultaneously, it is necessary to understand their respective frequency-dependent contributions to the THz detector signal in order to design and realize the next generation high-performance G-TeraFET sensors. \textit{In this Work}, we address this issue and present a THz characterization of CVD G-TeraFETs by using a methodology that reveals RSM and PTE response contributions independently. This is achieved by comparing photoresponses obtained with symmetric antenna coupling (SAC) and asymmetric antenna coupling (AAC) schemes. Our graphene devices were fabricated by a wafer-scale process of Ref. \onlinecite{Generalov2022}, which was developed to improve the reproducibility and scalability of graphene device fabrication, thus allowing reliable comparison of neighboring devices with different designs. We show that the detector sensitivity of the G-TeraFET devices with the SAC design is dominated by the PTE and that their performance is comparable to that of devices with the AAC layout. \textcolor{black}{For the AAC design,  we find that the contribution of the RSM to the measured detector signal decreases dramatically with increasing frequency compared to the contribution of the PTE; the ratio is 1:1 at 100~GHz and 1:5 at 1000~GHz.} 

\section{Identification of the rectification mechanisms by device geometry}
\begin{figure*}[t!]
\centering
\includegraphics[keepaspectratio, width=1\textwidth]{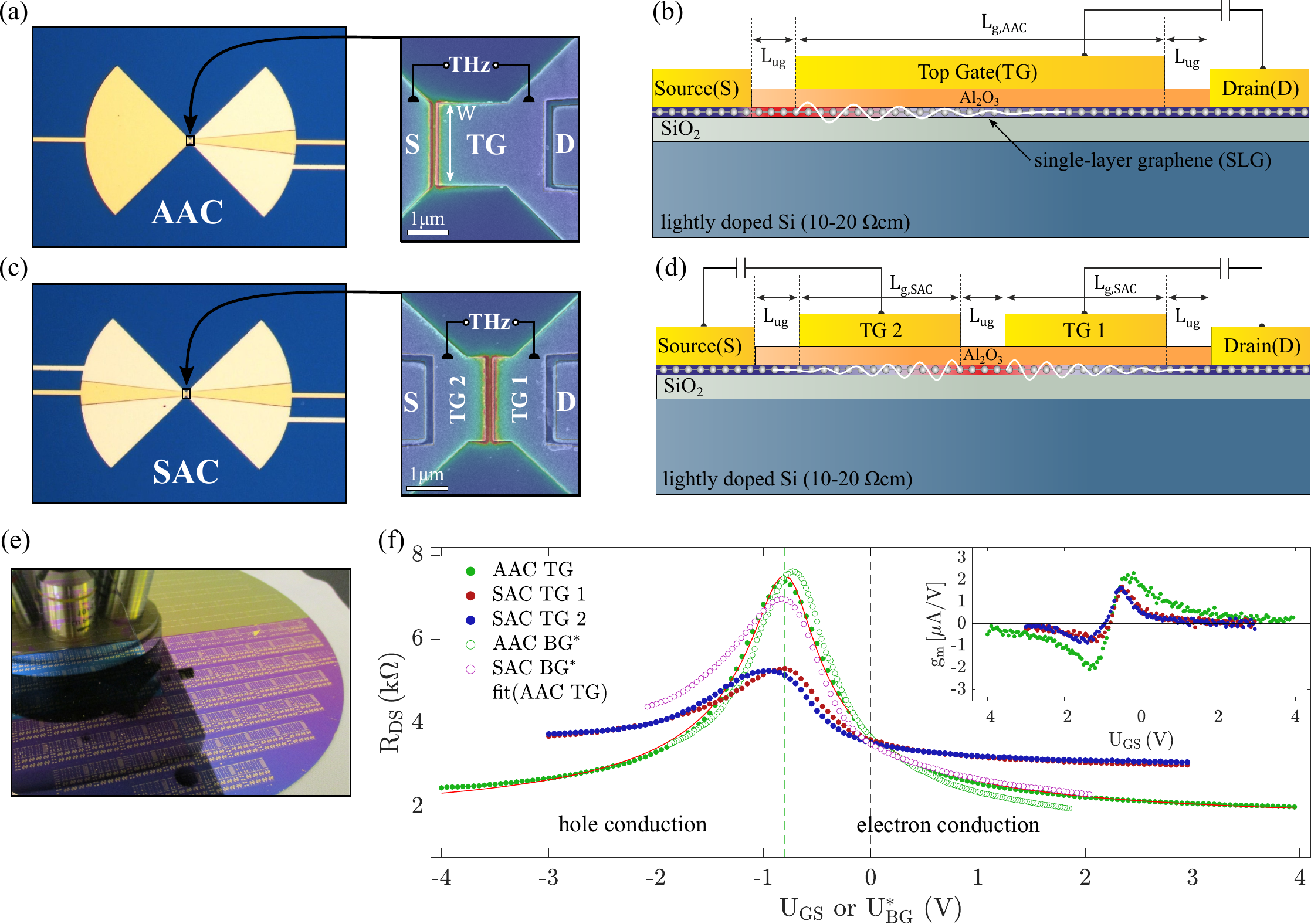}
\caption{\textbf{(a,b) Asymmetric Antenna Coupling (AAC).} (a) Micrograph of the single-split bow-tie geometry and magnified image of its central region obtained by scanning electron microscopy (SEM). ``S'' denotes source, ``D'' denotes drain, and ``TG'' denotes top gate. The color-shaded overlay on the SEM image represents the simulated THz electric field distribution at 400~GHz calculated with the Maxwell solver CST Studio Suite. (b) Cross-sectional schematic view of the central region of the device to illustrate the PTE and RSM rectification processes. The red color superimposed on the graphene sheet (indicated by the white line and dots) shows where the THz field collected by the antenna heats the charge carriers in the graphene and induces a PTE. The white waves represent plasma waves induced by the THz field. They are associated with RSM.
\textbf{(c,d) Symmetric Antenna Coupling (SAC).} (c) Micrograph of the detector with double-split bow-tie antenna and SEM image of the channel region with a color-coded overlay showing the THz field amplitude at 400~GHz. (d) Schematic illustrating the rectification processes. (e) Image of a 6" Si/SiO$_2$ wafer with finished TeraFETs. (f) Measured drain-to-source resistance $R_{DS}$ of an AAC and a SAC device. The legend indicates which gate voltage is tuned (TG or the BG in the case of the AAC device, or either one of TG~1, TG~2 or BG in the case of the SAC device). Open green dots represent BG measurements, the curves are plotted against the scaled BG voltage  ${U_{BG}^*} = U_{BG} \cdot \left( C_{ox,BG}/ C_{ox,TG} \right)$). Inset: Measured transconductances $g_m$ as a function of the TG voltages.}
\label{fig:schematics+IV}
\end{figure*}

To distinguish the PTE and the RSM contributions in the photoresponse of G-TeraFETs, we designed and fabricated detectors with two different antenna coupling schemes, AAC (\Figrefa{fig:schematics+IV}) and SAC (\Figrefc{fig:schematics+IV}), following the approach of \onlinecite{Viti2021}. Both types of detectors use the same bowtie antenna \textcolor{black}{with a radius of $120~\mu$m} coupled to a single- or double-gated graphene channel \textcolor{black}{with a length of $2.7~\mu$m and a width of $2~\mu$m} at the center of the antenna.
In the AAC design one antenna leaf leads to the S electrode of the graphene FET, and the other -- in a split antenna configuration -- to the top gate (TG) and D electrodes, see \Figrefa{fig:schematics+IV} and \Figrefb{fig:schematics+IV}. The split results in capacitive coupling between the D and TG leaves of the antenna, shunting them to the same AC potential. As a result of the capacitive shunting, the THz electric field oscillates between the S and TG electrodes (see color-coded overlay of the field amplitude on the SEM image in \Figrefb{fig:schematics+IV}). 
The oscillating THz field leads to (i) carrier heating and a PTE signal (indicated by the color shading in \Figrefb{fig:schematics+IV}), and (ii) excitation of plasma waves (indicated by the white wavy line), accompanied by the build-up of a rectified voltage due to the RSM. 

The SAC device is shown in \Figrefc{fig:schematics+IV} and \Figrefd{fig:schematics+IV} and is optimized for photothermoelectric effect rectification. The SAC antenna layout has two top gate (TG) electrodes -- TG~1 and TG~2 -- to tune the conductivity of the two gated regions of the graphene channel separately and create a p-n junction. The THz electric field is mainly confined to the central channel region between the two top gates (see plot of the simulated field distribution superimposed on the SEM image of the device in \Figrefc{fig:schematics+IV} \textcolor{black}{and Fig. S6(c),(d) in the Supporting Information}). Both antenna leaves have splits that shunt the TG~1 with D and TG~2 with S. This minimizes out the RSM response in SAC (the plasma waves propagate in opposite directions and the rectified voltages are of opposite sign) G-TeraFETs, while keeping the antenna geometries and channel dimensions for both designs are as similar as possible. 
The width $W$ and the total length $L$ of the channel (distance between S and D) are the same for SAC and AAC, being $W=2~\mu$m and $L=2.9~\mu$m. For AAC, according to \Figrefb{fig:schematics+IV}, $L = L_{AAC} = L_{g,AAC}+2 \cdot L_{ug}$, with $L_{g,AAC}=2.5~\mu$m and $L_{ug}=0.2~\mu$m. For SAC, as shown in \Figrefd{fig:schematics+IV}, $L=L_{SAC}=2\cdot L_{g,SAC}+3\cdot L_{ug} = L_{AAC}$, with $L_{g,SAC}=1.15~\mu$m. We fabricated additional AAC-type devices with different values of $L_{ug}$ ($L_{ug}= 0.1,\,0.2, \,0.3~\mu$m) in order to study the influence of the ungated segments of the channel (for results see  \Figrefa{fig:RSMvsPTE}). We optimize the responsivity of AAC- and SAC-type devices by using a high resistivity (undoped) Si substrate, thus reducing the absorption losses in the substrate (see Fig.~\ref{fig:SAC_PTE_MAP} and Fig.~\ref{fig:NEP}, and Discussion).

In the following, the two rectification phenomena are discussed in more detail for each detector design, first PTE and then RSM. 
Electromagnetic wave (EM) simulations with CST Studio Suite identify the regions, where the THz field drives a high-frequency current and thus heats the carrier ensemble in the graphene sheet.  
The inhomogeneity of the THz field leads to spatial gradients in the carrier temperature $T_C$ as indicated by the red (hot, $T_C>T_L$) to blue (cold, $T_C=T_L$) color-coded overlay on the plots of the graphene sheet in \Figrefb{fig:schematics+IV} and \Figrefd{fig:schematics+IV}.
The temperature-dependent change in local Fermi energy causes the carriers to diffuse away from the hot region. \\
In the AAC case, the graphene sheet together with the S and D electrodes forms a thermocouple with two junctions at the graphene-S and graphene-D interfaces, where the carriers of the first junction are heated, while the second junction remains cold. Under steady-state conditions, the diffusion current leads to the formation of a PTE voltage $U_{PTE}$ (Seebeck effect) between D and S, which under open circuit conditions can be approximated by \cite{Viti2021,Bandurin2018b}
\begin{equation}
\begin{split}
    U_{PTE} & = {\int_S}^D \frac{\partial T_C(x)}{\partial x}  S(x)dx \\
            & \approx \Delta T_C \cdot\left(S_{ug}^{hot} - S_g(U_{GS})\right),
    \label{eq:PTE}
\end{split}
\end{equation}
where $\Delta T_C = (T_{ug}^{hot}-T_g)$, with $T_g = T_L \; (\approx \rm{293~K})$. Here, the heated ungated region and the remaining part of the transistor channel are considered as two lumped elements. 
$S_g(U_{GS})$, the Seebeck coefficient (thermopower) of the gated graphene channel, can be controlled by the TG voltage $U_{GS}$, while the Seebeck coefficient of the ungated segment of the graphene channel has a fixed value $S_{ug}\approx S_g(0\,V)$. It is important to note, that the Seebeck coefficients and thus $U_{PTE}$ also depend on the back gate (BG) voltage $U_{BG}$, which is not explicitly expressed in the equation. The same is true for the conductivity and the Fermi energy in the following equations. 
The Seebeck coefficient $S_g(U_{GS})$ can be calculated from the semiclassical Mott formula for degenerate semiconductors according to \cite{Viti2021,Bandurin2018b,Ghahari2016,Jonson1980,Mott1969}
\begin{align}
 S_g(U_{GS}) = - \frac{\pi^2 k_B T_L}{3e} \frac{\partial  \ln (\sigma(U_{GS}))}{\partial \, U_{GS}} \frac{\partial \,U_{GS}}{\partial \, E_F}\,,
 \label{eq:Mott}
\end{align}
where $e$ is the elementary charge, $\sigma(U_{GS})$ the electrical conductivity of the channel, and $E_F$ the Fermi energy in the gated region of the graphene sheet. It is difficult to accurately determine $\sigma(U_{GS})$ for a given device. We obtain an approximate $U_{GS}$-dependent value from the measured drain-to-source resistance $R_{DS}(U_{GS})$ via $\sigma(U_{GS}) = (R_{DS}(U_{GS}))^{-1} \cdot (W/L)$. $R_{DS}$ includes the resistance contribution of the ungated regions and the contact resistances. 
$E_F$ in Eq.~(\ref{eq:Mott}) is calculated by \cite{Ghahari2016}
\begin{align}
\begin{split}
E_F & = \hbar v_F \sqrt{\pi  C_{ox,TG} \, |U_{GS}-U_{Dirac}|/e} \\ 
& \hspace{4mm} \cdot \sgn (U_{GS}-U_{Dirac}).
\end{split}
 \label{eq:Ef}
\end{align}
Here, $v_F$ is the SLG Fermi velocity and $U_{Dirac}$ the Dirac voltage. $C_{ox,TG}$ denotes the capacitance per unit area of the TG. 
\begin{figure*}[!t]
\centering
\includegraphics[keepaspectratio, width=1\textwidth]{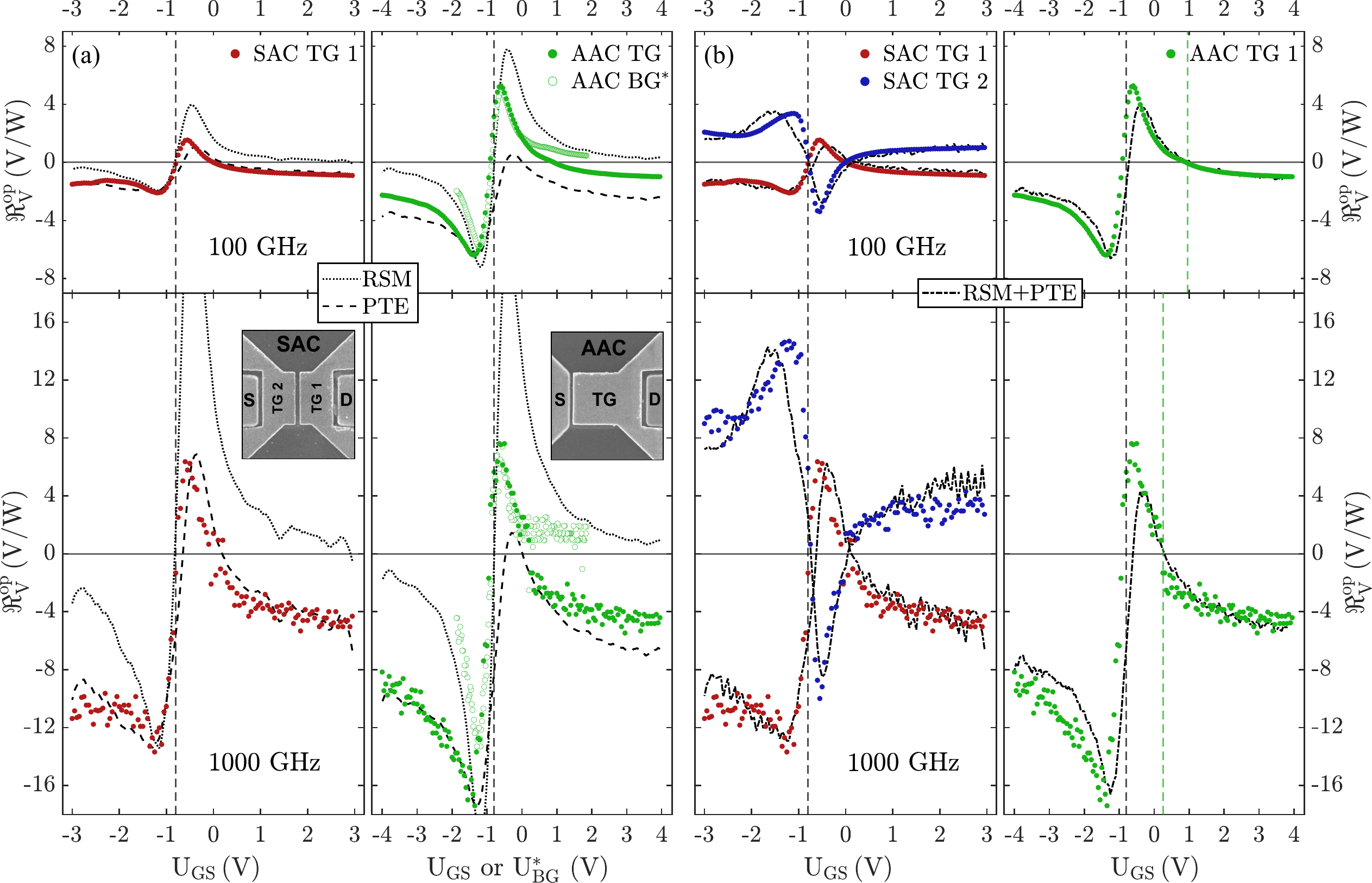}
\caption{(a) Measured optical voltage responsivity ${\Re_V^{op}}$ of G-TeraFETs with symmetric antenna coupling (SAC, left panel, red) and asymmetric antenna coupling (AAC, right panel, green). The calculated photoresponse for resistive self-mixing (RSM, dotted lines, calculated with Eq.~(\ref{eq:Sako})) and photothermoelectric rectification (PTE, dashed lines, calculated with Eq.~(\ref{eq:PTE}) together with Eq.~(\ref{eq:Mott})) are also shown. (b) Measured and calculated optical responsivities. The calculated curves were obtained by fitting a linear combination of the theoretical RSM and PTE photoresponses (RSM+PTE, dash-dotted lines) to the measured curves. In the case of AAC, the second sign change of the photoresponse is indicated by a vertical green dashed line.}
\label{fig:resp}
\end{figure*} 
In the SAC case a pair of thermocouples is formed, i.e., a thermopile with two segments, one on each side of the graphene channel. 
On the right side (TG~1 and D), one thermocouple consists of the heated junction between the central ungated graphene and the cold junction at the D contact side TG1--graphene--D, similar to the AAC case. The second thermocouple, on the left side (TG~2 and S) of the graphene channel, is formed between the cold junction of S--graphene--TG2 and the heated junction between TG~2 and the central ungated graphene. The two thermocouples are oriented opposite to each other. One gets a net diffusion current and $U_{PTE} \neq 0$~V only when TG~1 and TG~2 are differently biased. Under unequal TG bias conditions, the resulting thermovoltage of the SAC thermopile is given by
\begin{equation}
    U_{PTE}^{SAC} \approx \Delta T_C \cdot\left(S_{TG2} - S_{TG1}\right),
    \label{eq:PTE-SAC}
\end{equation}
where $S_{TG2}$ and $S_{TG1}$ represent the TG voltage dependent Seebeck coefficients in each gated graphene region (TG~2 and TG~1). If TG~2 is set to ground potential ($S_{TG2} = S_{ug}$), thus cancelling the thermoelement on the left (S) side, one retrieves the AAC case with $U_{PTE}^{TG1} \approx \Delta T_C \cdot\left(S_{ug}^{hot} - S_{TG1}\right)$. Consequently, when TG~1 is set to ground potential, the expected thermovoltage of the SAC photodetector can be calculated from $U_{PTE}^{TG2} \approx \Delta T_C \cdot\left(S_{TG2} - S_{ug}^{hot}\right)$.

We now turn to the RSM effect. It is present in a single-split bow-tie geometry of the AAC layout because the THz radiation is asymmetrically coupled to the transistor channel. Under such conditions, rectification occurs by resistive mixing at any frequency of the THz wave -- at sufficiently high frequency influenced by a plasma wave which is excited in the channel as shown in \Figrefb{fig:schematics+IV}\cite{Boppel2012,Dyakonov1996}. The rectified RSM voltage as a function of the TG voltage $U_{GS}$ can be estimated by\cite{Sakowicz2011,Vicarelli2012}
\begin{equation}
    U_{RSM} = \left(\frac{{U^{el}_{THz}}}{2}\right)^2 \cdot \frac{\partial \, {\ln}(\sigma(U_{GS}))}{\partial \, U_{GS}},
    \label{eq:Sako}
\end{equation}
where ${U^{el}_{THz}}$ is the voltage amplitude of the THz signal between S and TG. 
Eq.~(\ref{eq:Sako}) is valid under two conditions: (i) the long-channel condition must hold, which is fulfilled here\cite{Soltani2020} ($s\cdot \tau_p < L_{g,AAC}$, where $s$ is the phase velocity of the plasma wave and $\tau_p$ the decay time of the plasma wave, which is identical to the momentum relaxation time of the charge carriers), and (ii) the radiation frequency must be in the regime of overdamped plasma oscillations ($\omega \cdot \tau_p < 1$, with $\omega=2\pi f_{THz}$, where nonlinear self-mixing effects of the induced plasma waves are not yet active \cite{Bauer2019,Boppel2012,Lisauskas2009}). For the frequencies of our measurements ($f_{THz}$ up to 1.2~THz), the 2nd condition is also fulfilled, since $\tau_p \approx$ 60~fs \cite{Soltani2020}. \textcolor{black}{Note that the performance of the G-TeraFET due to the rectification by resistive self-mixing can also be modeled by means of Volterra series approaches\cite{Andersson2016,Yang2020}.} \\
In case of the SAC design, no net RSM signal is obtained as long as the two top gates are equally biased (relative to $U_{Dirac}$). If this is not the case, \textcolor{black}{one} can employ Eq.~(\ref{eq:Sako}) for each side of the central ungated region and calculate the net $U_{RSM}$ value. 

\section{DC and THz characterization}
\subsection{DC transport properties}

In \Figreff{fig:schematics+IV}, we present the DC electrical transport characteristics -- the drain-to-source resistance $R_{DS} =  U_{DS} / I_{DS}$ and the transconductance $g_m = \Delta I_{DS}/ \Delta U_{GS}$ (shown in the inset) -- as a function of either the TG voltage $U_{GS}$ or the back gate voltage $U_{BG}$ for a device of AAC or SAC design, respectively (for device fabrication, see Methods section). During the recording of each curve, only one electrode was biased as specified in the legend of the graph -- either TG or BG in the case of the AAC device, or TG~1, TG~2 or BG in the SAC case, while the other gate electrode(s) was (were) set to ground potential.
For ease of comparison, we plot the curves obtained by varying the BG voltage, as a function of a scaled voltage  
${U_{BG}^*}= U_{BG} \cdot (C_{ox,BG}/C_{ox,TG})$, where $C_{ox,TG}$ and $C_{ox,BG}$ represent the TG- and BG-to-channel capacitances per unit area, respectively. 
All devices on the wafer exhibit similar electrical transport characteristics -- Dirac voltages $U_{Dirac}$ close to -0.8~V, peak channel resistances around 7~k$\Omega$ -- for both device designs (AAC and SAC). We attribute this to the high homogeneity and reproducibility of the wafer-scale fabrication process including the global transfer of CVD graphene to the entire wafer.

The field-effect mobilities $\mu_{FE,e}$ and $\mu_{FE,h}$ for electrons and holes, the contact resistance $R_C$ and the residual carrier concentration $n_0$ of the AAC device can be obtained from the measured $R_{DS}(U_{GS})$ curve using Eq.~(\ref{eq:Rfit_model}) given in the Methods section as a fit function. 
\Figreff{fig:schematics+IV} shows a fit of the dependence of Eq.~(\ref{eq:Rfit_model}) to the measured TG $R_{DS}$ curve of AAC using the parameter values $L_{g,AAC}=2.5~\mu$m, $L_{ug}=0.2~\mu$m, $C_{ox,TG}=0.37~\mu$F$\,$cm$^{-2}$, $U_{Dirac}= -0.8$~V. One obtains from the fitting: $\mu_{FE,e} = 1631$~cm$^2$/Vs, $\mu_{FE,h} = 1257$~cm$^2$/Vs, $R_{C} = 1205~\Omega$, $n_{0} = 4.54 \cdot 10^{11}$~cm$^{-2}$. These values are consistent with the values reported for a CVD G-TeraFET using a similiar device layout in \onlinecite{Zak2014}. 

\subsection{Frequency dependency of the responsivity}
Following the DC electrical characterization, each TeraFET was characterized in the THz frequency range 0.1-1.2~THz. 
The \textit{optical} voltage responsivity $\Re_V^{op} = \Delta U_{DS} / P^{op}_{THz}$ was determined from the rectified voltage $\Delta U_{DS}$ and the total THz beam power $P^{op}_{THz}$ measured in steps of 100~GHz as a function of the bias voltage applied to the respective gate. 
Results are shown in \Figref{fig:resp}. In the case of the AAC device, this was either TG or BG, in the case of the SAC device either TG~1 or TG~2, while the other gate electrode was set on ground potential.

In \Figrefa{fig:resp}, we present $\Re_V^{op}$ at 100 and 1000~GHz for the SAC- and AAC-type devices which were also used for the measurements of \Figreff{fig:schematics+IV}. The vertical dashed lines mark the Dirac voltage $U_{Dirac}$. 
\Figrefa{fig:resp} also shows theoretical predictions for the photoresponse based on the RSM mechanism and the PTE mechanism. 
The unknown prefactors of Eq.~(\ref{eq:PTE}) and Eq.~(\ref{eq:Sako}), indexed by 0 here -- $\Delta T_{C,0}$ in case of PTE and (${U^{el}_{THz,0}})^2$ in case of RSM --, are fixed by letting the calculated curves go through the point with the largest magnitude of the detected signal (for these devices, this is for negative values of $\Re_V^{op}$ observed at the hole conduction branch -- left of $U_{Dirac}$ -- for values of $U_{GS}$ close to -1~V). 
When we consider the TG~1 SAC photoresponse, one finds that the measured curves are better approximated by the PTE model than the RSM model. The same holds for the TG AAC photoresponse at 1000~GHz, but not at 100~GHz.
For BG biasing we observe a deviation from RSM-type rectification, which is is especially visible on the electron-conduction side of the gate bias (positive $U_{BG}^{\star}-U_{Dirac}$). We explain the deviation by a PTE contribution to the rectified voltage. For PTE to contribute, there must be an asymmetry in the Seebeck coefficients. Indeed, the literature reports that the Fermi level of graphene is tuned differently by a BG depending on whether there is an additional metal layer on top of the graphene or not\cite{Chaves2015}. \textcolor{black}{The metal pins the Fermi level of graphene, leading to contact doping in the vicinity of the metal contact. It has been shown by near-field scanning optical microscopy measurements that this contact doping effect extends 0.2 to 0.3 $\mu$m into the graphene channel\cite{Müller2009}. This suggests that there is a Seebeck difference even for BG operation, which leads to an asymmetric carrier diffusion in graphene and a concomitant build-up of a PTE voltage.} \\
\begin{figure*}[!t]
\centering
\includegraphics[keepaspectratio, width=1\textwidth]{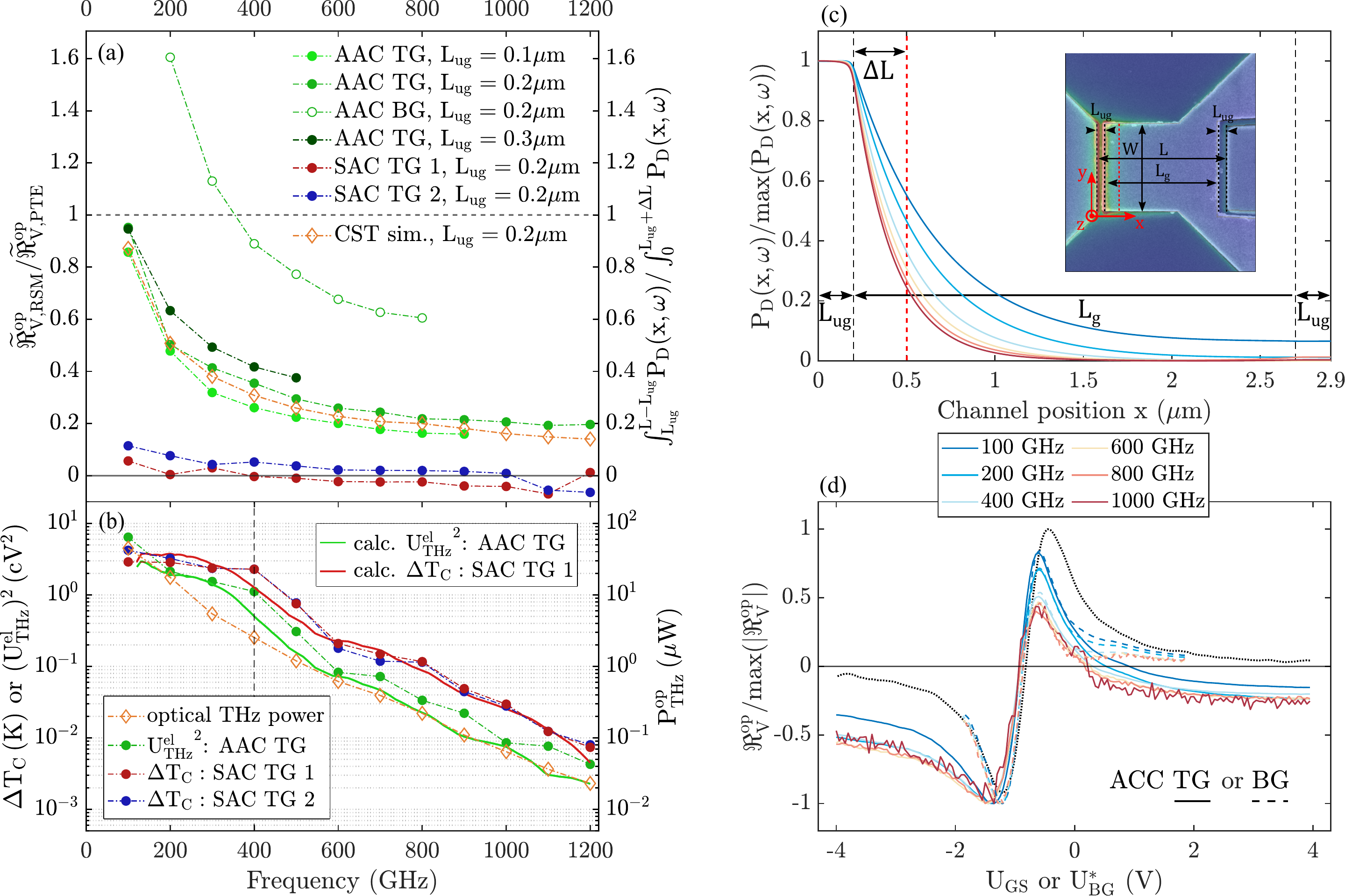}
\caption{(a) Ratio of the extracted responsivities ${\widetilde{\Re}}_{V,RSM}^{op}/{\widetilde{\Re}}_{V,PTE}^{op}$ (left vertical axis) for AAC-type devices (green) and a SAC-type one (red for biasing via TG~1 and blue for TG~2). For comparison: CST-simulated ratio of the total surface power loss density attributed to RSM vs. that attributed to PTE (yellow, right vertical axis; see main text for details).  
(b) Frequency dependence of the calculated RSM and PTE signals, expressed by the respective prefactors $({U^{el}_{THz}})^2$ \textcolor{black}{(full green dots, unit in centi-volt (cV) squared)} and $\Delta T_C$ (full red dots for biasing via TG~1 and full blue dots for TG~2). % response models obtained from the RSM+PTE fitting routine. 
Open yellow diamonds: Power $P^{op}_{THz}$ of the THz beam measured at the detector position. \textcolor{black}{Solid lines: Estimated prefactors $({U^{el}_{THz}})^2$ for the RSM signal of AAC TG (green) and $\Delta T_C$ for the PTE signal of SAC TG 1 (red) determined from $P^{op}_{THz}$ and antenna simulations.}
(c) Calculated normalized surface power loss density $P_D (x,\omega)$ along the graphene channel ($x$-direction) as a function of THz frequency for a AAC-type device with $L_g=L_{g,AAC}=2.5~\mu$m and $L_{ug}=0.2~\mu$m. The calculation was performed with the CST Studio Suite. The inset depicts a SEM picture of the AAC detector's channel region overlayed by the simulated field distribution at 400~GHz and the coordinate system (shown in red). 
The red dashed vertical line (at $x= 0.5~\mu$m, $\Delta L=0.3~\mu$m) marks the effective PTE penetration \textcolor{black}{length} in the gated channel region. (d) Measured normalized detector responsivity for the AAC coupling scheme in the case of TG tuning (solid lines) and BG tuning (dashed lines) and the calculated RSM photoresponse using Eq.~\ref{eq:Sako} (dotted line) for different THz frequencies.}
\label{fig:RSMvsPTE}
\end{figure*} 
Coming back to the fitting of all $\Re_V^{op}$ curves, we now address the question whether a combination of both rectification mechanisms better describes the measured photoresponse than each single effect. We follow an approach similar to that in \onlinecite{Viti2020} and express the total photoresponse signal $\Delta U_{DS}$ as a linear combination of calculated RSM and PTE rectification contributions:
\begin{equation}
\begin{split}
    \Delta U_{DS,c.} & (U_{GS,BG},\omega) = \\ 
    & X_{RSM}(\omega) \cdot U_{RSM}(U_{GS,BG},\omega) \\
    & + X_{PTE}(\omega) \cdot U_{PTE}(U_{GS,BG},\omega),
    \label{eq:Linear_model}
    \end{split}
\end{equation}
where the frequency-dependent, but voltage-independent fit parameters $X_{RSM}(\omega)$ and $X_{PTE}(\omega)$ are obtained by least-mean-square fits to the measured photoresponse. The prefactors $\Delta T_{C,0}$ and $(U^{el}_{THz,0})^2$ of $U_{RSM}$ and $U_{PTE}$ are those of \Figrefa{fig:resp}.  \\
In \Figrefb{fig:resp}, we present measured $\Re_V^{op}$ curves with TG voltage tuning (those of \Figrefa{fig:resp} plus additional ones for SAC with biasing of TG~2), and overlay them with the best fit curves (RSM+PTE). \\
They were calculated by
\begin{equation}
\begin{split}
    \Re_{V,c.}^{op}(U_{GS,BG},\omega) & = \frac{\Delta U_{DS,c.}}{P^{op}_{THz}} \\
     & = \Re_{V,RSM}^{op} + \Re_{V,PTE}^{op}, 
    \label{eq:Linear_model_Resp}
\end{split}
\end{equation}
where we introduced the intrinsic calculated responsivities of the RSM and the PTE rectification processes via
\begin{equation}
\begin{split}
    \Re_{V,RSM}^{op}(U_{GS,BG},\omega) = X_{RSM} \cdot U_{RSM}/P^{op}_{THz} \\
    \Re_{V,PTE}^{op}(U_{GS,BG},\omega) = X_{PTE} \cdot U_{PTE}/P^{op}_{THz}. \\
    \end{split}
    \label{eq:Resp_RSM_PTE}
\end{equation}
We observe a near-quantitative agreement between the RSM+PTE curves and the measured photoresponses for both detector layouts, and clearly a significantly better agreement than is obtained with each rectification effect alone in \Figrefa{fig:resp}. For the AAC device, even the second sign change of the photoresponse -- besides the sign change at the Dirac point -- %($U_{Dirac}$, shown as vertical black dashed line) 
is correctly reproduced by the RSM+PTE curves. The good agreement implies that the total THz photoresponse of G-TeraFETs can with good confidence be broken down to a linear combination of RSM and PTE. And apparently both models are valid over the covered range of bias voltages and THz frequencies. \\
We have repeated this kind of analysis for a number of frequencies (from 0.1 to 1.2~THz in steps of 100~GHz) and for several devices including AAC-type devices with various values of the ungated channel length $L_{ug}$ ($L_{ug} = 0.1, \, 0.2,\, 0.3~\mu$m). Results for several AAC and SAC photodetectors are shown in \Figrefa{fig:RSMvsPTE}. We plot the frequency-dependent RSM-to-PTE responsivity ratio ${\widetilde{\Re}}_{V,RSM}^{op}(\omega)/{\widetilde{\Re}}_{V,PTE}^{op}(\omega)$. The tilde accent stands here for a special choice of the gate voltage, which was applied either to TG, BG, TG~1 or TG~2. For each device and choice of gate electrode, we first determined the gate voltage which yielded the highest responsivity value at 400~GHz. This gate voltage was then maintained also for all other frequencies. \\
In \Figrefa{fig:RSMvsPTE} it is clearly seen that all AAC photodetectors exhibit a similar roll-off of ${\widetilde{\Re}}_{V,RSM}^{op}/{\widetilde{\Re}}_{V,PTE}^{op}$ with frequency. At 100~GHz, RSM and PTE are of similar strength, but with increasing frequency, PTE becomes more and more dominant. 
The behavior is weakly dependent on the value of $L_{ug}$, as one finds when considering the ratio
\begin{align*}
\frac{{\widetilde{\Re}}_{V,RSM}^{op}\rm{(100~GHz)}/{\widetilde{\Re}}_{V,PTE}^{op}\rm{(100~GHz)}}{{\widetilde{\Re}}_{V,RSM}^{op}\rm{(500~GHz)}/{\widetilde{\Re}}_{V,PTE}^{op}\rm{(500~GHz)}} \;, 
\end{align*} 
which yields a value of 2.5 for $L_{ug} = 0.3~\mu$m, 3.2 for 0.2~$\mu$m and of 3.8 for 0.1~$\mu$m. \textcolor{black}{Interestingly, we also observe such a roll-off with frequency for the BG measurement, which we attribute to the contact doping effect \cite{Müller2009,Vangelidis2022} that leads to the metal-contacted SLG/SLG PTE discussed above. Without this doping effect, one would expect a dominant RSM over the PTE, since the Seebeck difference between the ungated and gated channel regions is not altered when a BG voltage is applied globally to the entire graphene channel (see also Supporting Information, S4). Note that the metal-contacted SLG/SLG PTE effect is particularly pronounced in our AAC-type detectors as our heated ungated channel gap length ($L_{ug}\approx0.1-0.3\,\mu$m) is of same order as the length scale of the metal contact doping effect\cite{Müller2009}. We have found further experimental evidence that the Seebeck coefficient of the ungated channel region $S_{ug}^{hot}$ in our AAC-type detectors is partially influenced by this contact doping effect. For further details we refer the reader to the Supporting Information (sections S4 and S8).}

For the SAC-type device, the ratio ${\widetilde{\Re}}_{V,RSM}^{op}(\omega)/{\widetilde{\Re}}_{V,PTE}^{op}(\omega)$ is always close to zero, as expected. In the SAC layout, the RSM contribution, compared with the PTE one, is significantly reduced by design, as discussed above. Note that for higher frequencies, the extracted responsivity ratio turns to negative values due to negative values of $X_{RSM}$ at these frequencies. Here, the least-mean-square fit routine suggest a negative RSM contribution to the detected signal of SAC, which may arise whenever the net RSM signal between TG~1 and TG~2 is of opposite sign with regard to the PTE signal. 

We now investigate the frequency dependence of the rectified voltages of the two mechanisms, and how they scale with the frequency dependence of the power of the radiation source used. 
The frequency dependence of the RSM contribution to the rectified voltage is contained in the product $({U^{el}_{THz}})^2 = X_{RSM} \cdot ({U^{el}_{THz,0}})^2 $, while that of the PTE contribution is in the product $\Delta T_{C} = X_{PTE} \cdot \Delta T_{C,0}$. 
\Figrefb{fig:RSMvsPTE} displays these two quantities as a function of frequency. For comparison, we also plot the frequency dependence of the power $P^{op}_{THz}$ of the THz beam. The power roll-off is typical for the radiation source, a Toptica TeraScan 1550 spectroscopy system. Comparing the slopes of the power roll-off and of the prefactors, one finds a good agreement -- except below 400~GHz (marked by vertical black dashed line). We attribute this to a poor impedance matching between the detectors' antennae and transistor channels below 400~GHz, which becomes worse with decreasing frequency (see \Figrefa{fig:NEP}). \textcolor{black}{Next, we check whether the observed frequency dependence and the overall magnitude of the extracted prefactors are reasonable for the given incident THz power $P^{op}_{THz}$. For this purpose we calculated $({U^{el}_{THz}})^2$ for AAC TG and $\Delta T_{C}$ for SAC TG 1 based on antenna simulations of each detector layout. Optical losses due to substrate absorption and reflection at the silicon lens as well as the impedance mismatch between the antenna (cp. \Figrefa{fig:NEP}) and the graphene channel are considered for the calculation. The increase of the carrier temperature ($\Delta T_{C}$) due to the THz power dissipated in the antenna gap is estimated from the steady-state carrier heating equation\cite{Massicotte2021}. For further details the reader is referred to the Supporting Information (section S9). We find semi-quantitative agreement in magnitude and roll-off behavior between the theoretical calculations and the extracted prefactors, suggesting that the fitting approach used in this work provides reasonable physical orders of magnitude for the prefactors of both rectification processes.} 

\subsection{THz power dissipation - origin of the dominance of PTE}
In order to better understand the origin of the observed roll-off of the ${\widetilde{\Re}}_{V,RSM}^{op}/{\widetilde{\Re}}_{V,PTE}^{op}$ ratio of the AAC-type devices, we performed EM wave simulations (CST Studio Suite) of the full antenna structure with S, D and G electrodes and a graphene layer buried under a 20-nm-thick Al$_{2}$O$_{3}$ insulator layer. 

Next, we determined the power dissipated in the graphene layer at the respective channel position using the surface loss density (SLD) function of CST (for further details, see Methods section).
In \Figrefc{fig:RSMvsPTE}, we plot the normalized SLD $P_D(x,\omega)/\rm{max}(P_D(x,\omega))$ for an AAC-coupled device with $L_{ug} = 0.2~\mu$m and $L_{g} = 2.5~\mu$m excited at various THz frequencies. $x$ is the coordinate along the transistor's channel, with $x=0$ marking the transition from S electrode to the ungated graphene (see \Figrefc{fig:RSMvsPTE}). The SLD is highest in the ungated graphene, and steadily decreases when entering the gated graphene from S side and proceeding towards D electrode. 
Remarkably, one observes a faster decrease of the SLD along the gated graphene when the THz frequency is increased. Since the SLD represents the loss of power of the THz plasma wave in the graphene, the faster SLD decay stands for a faster decay of the plasma wave. 

We now utilize the information about $P_D(x,\omega)$ to estimate how strong the rectification by PTE should be relative to that by RSM. In order to do so, we assume that the rectified voltage scales with the total power dissipated in the respective spatial regions where the rectification occurs. In case of the RSM mechanism, this is the gated part of the channel. With regard to PTE, rectification occurs in the ungated channel region and in the gated part of the channel down to an unknown effective \textcolor{black}{length} $\Delta L$. In what follows, we furthermore assume $\Delta L$ to be independent of $\omega$. Based on these suppositions, we first integrate the SLD along the gated channel region, and then divide the resulting quantity by the integral over the SLD from $x=0$ to $x=L_{ug} + \Delta L$. The resultant ratio should be directly proportional to the measured ${\widetilde{\Re}}_{V,RSM}^{op}/{\widetilde{\Re}}_{V,PTE}^{op}$ ratio if the assumption stated above is correct.

In \Figrefa{fig:RSMvsPTE} we plot the SLD-integral ratio as a function of frequency. One can directly compare the results with the measured ${\widetilde{\Re}}_{V,RSM}^{op}/{\widetilde{\Re}}_{V,PTE}^{op}$ ratio of the corresponding AAC-type device with $L_{ug} = 0.2~\mu$m and TG biasing. We find excellent agreement of the frequency roll-off of the two ratios if we choose $\Delta L = 0.3~\mu$m. This finding has three implications: (i) It supports the basic validity of the underlying assumptions. (ii) It yields an effective \textcolor{black}{length} of 0.3~$\mu$m, to which the gated channel region contributes to PTE. (iii) And most importantly, the increasing domination of PTE over RSM for rising radiation frequency is a consequence of the \textcolor{black}{change in the power dissipation profile alongside the graphene channel when the frequency increases}. \textcolor{black}{The penetration length $\Delta L$ to which the PTE contributes to the detector signal below the gate electrode is closely related to the electronic cooling length $L_c$ \cite{Song2012, Antidormi2021,Vangelidis2022}, which scales sublinearly with the charge carrier mobility, and determines the length scale over which the heated carriers (here the carrier heating mainly occurs place in the antenna gap) cool down to the lattice temperature $T_L$. For our low carrier mobility samples, where the mobility is ranging from $\mu_{FE,h} \approx 1250$~cm$^2$/Vs up to $\mu_{FE,e} \approx 1600$~cm$^2$/Vs, an electronic cooling length of $L_{c,h} \approx 0.35\,\mu$m and $L_{c,e} \approx 0.38\,\mu$m is predicted from theory \cite{Antidormi2021}. This corresponds roughly to our assumption for $\Delta L$.} \textcolor{black}{To investigate the influence of carrier mobility on the RSM-to-PTE ratio of the G-TeraFET, we performed further EM wave simulations assuming different momentum relaxation times $\tau_p$, ranging from 60~fs (CVD-grown graphene) up to 300~fs (exfoliated graphene encapsulated in hBN). The simulations show that the higher the carrier mobility (higher values of $\tau_p$) the more THz power is dissipated in the gated transistor part (see Supporting Information Fig. S4), which could give rise to a larger RSM-to-PTE ratio in high mobility graphene THz detectors with AAC-type layout.} 

\begin{figure*}[!t]
\centering
\includegraphics[keepaspectratio, width=1\textwidth]{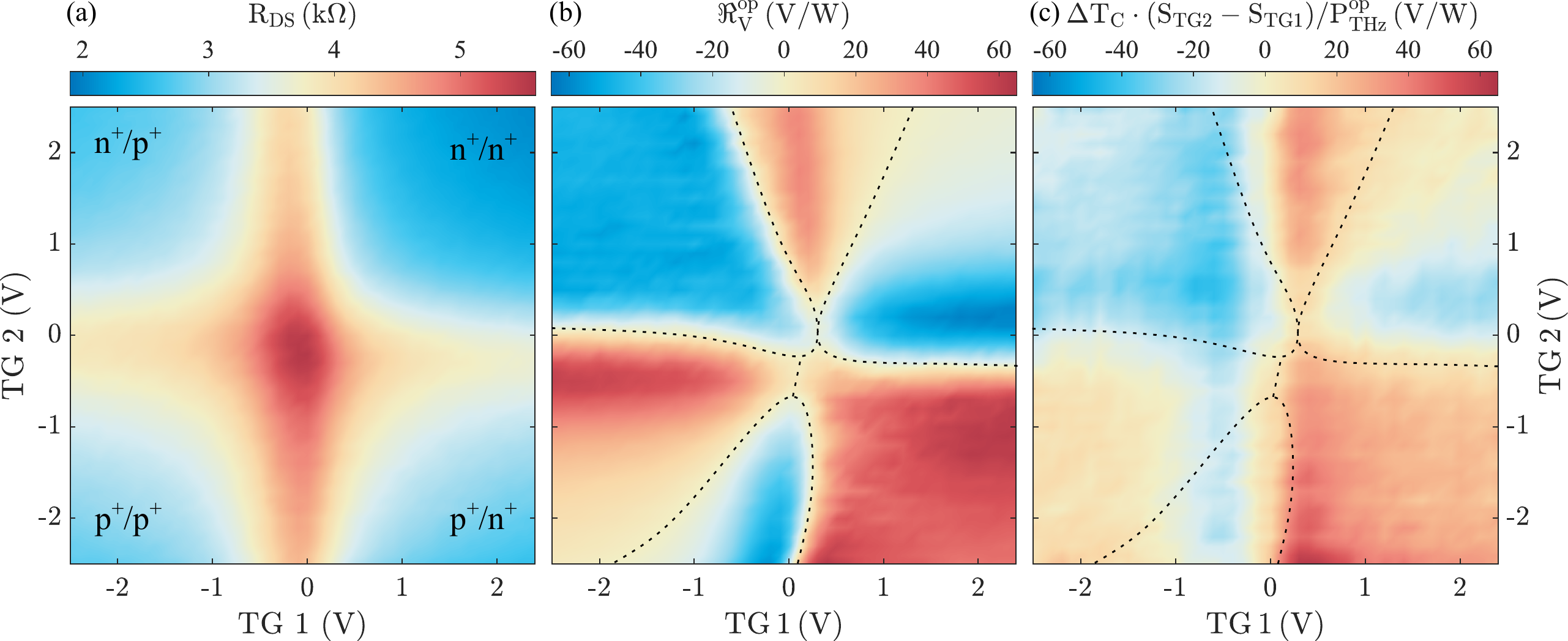}
\caption{(a) Measured drain-to-source resistance ${R_{DS}}$ and (b) optical voltage responsivity ${\Re_V^{op}}$ at 400~GHz of a G-TeraFET with SAC as a function of the gate voltages applied to both TG~1 and TG~2 (cp. \Figrefd{fig:schematics+IV}). (c) Calculated optical voltage responsivity at 400 GHz obtained from Eq.~(\ref{eq:PTE}). The black dashed lines (which are the same in (b) and (c)) mark the positions of the signal sign change in the measurements of (b).}
\label{fig:SAC_PTE_MAP}
\end{figure*}

Building on these computational findings, we want to add two observations here, which help to roughly estimate the relative importance of PTE, when inspecting a gate-voltage-dependent responsivity curve. \Figrefd{fig:RSMvsPTE} displays such curves for the AAC-type device with $L_{ug} = 0.2~\mu$m. Results are shown for several frequencies and for both TG and BG biasing. All curves are normalized to the peak absolute value of the respective responsivity curve. For the measured curves, this peak value always lies on the hole-conduction side of the gate voltage. The dotted black line is a calculated one for RSM detection alone for TG operation, obtained with Eq.~(\ref{eq:Sako}). The first observation which we want to point to, relates to the maximum of each normalized curve on the electron-conduction side. It is obvious that the value of this maximum decreases, if PTE becomes more important. The second observation relates to the zero crossing on the electron-conduction-side of the curves. For a pure RSM signal, one does not observe such a sign change. This is also the case for the measured curves with BG tuning, where the RSM contribution to the signal is large (see \Figrefa{fig:RSMvsPTE}). 
However all curves with TG biasing exhibit a sign change. The stronger the PTE is relative to RSM, the more the bias voltage, where the sign change occurs, shifts to lower values.

\subsection{Maximized thermoelectric detector response}
To reduce the THz losses through the substrate and maximize the detector performance the same detectors were manufactured on a high resistivity Si substrate with the specific resistance >10~k$\Omega$cm, compared to previously used Si substrate with the specific resistance of 10-20~$\Omega$cm which facilitate the BG functionality.
The results of the extensive characterization measurements performed on the SAC-type detector ($L_{g,SAC} = 1.2~\mu$m, $L_{ug} = 0.1~\mu$m) are shown in \Figref{fig:SAC_PTE_MAP}, where the bias voltages are applied to both to TG~1 and TG~2 independently from each other. 
The resulting 2D maps of $R_{DS}$ are shown in \Figrefa{fig:SAC_PTE_MAP}, and put into perspective with $\Re_V^{op}(f=400$~GHz) shown in \Figrefb{fig:SAC_PTE_MAP} and the calculated PTE contribution to the rectification displayed in \Figrefc{fig:SAC_PTE_MAP}.

In the color map of \Figrefa{fig:SAC_PTE_MAP}, one finds that $R_{DS}$ peaks at a value of 5.53~k$\Omega$ reached at $U_{GS}(\mathrm{TG1}) = 0$~V and $U_{GS}$(TG~2)$ = -0.2$~V close to the coordinate system's origin. 
Moving away from the origin, $R_{DS}$ declines, more slowly along the coordinate axes and faster into the four quadrants. The white text in the four quadrants indicates the type of conduction ($n$ or $p$), the first entry is for the graphene under TG~2, the second for the graphene under TG~1. \\
The color plot of \Figrefb{fig:SAC_PTE_MAP} shows $\Re_V^{op}$ measured at 400~GHz. Red and blue colors stand for the sign of the rectified voltage. 
The absolute value of $\Re_V^{op}$ is highest (>60~V/W) if the graphene segments under TG~1 and TG~2 form $\rm{n^+/p^+}$- or $\rm{p^+/n^+}$-junctions. In these case, the difference of the Seebeck coefficient of the two regions ($\Delta S = S_{TG2} - S_{TG1}$) is largest. The dashed lines marks the bias voltages where the rectified voltage and hence also $\Re_V^{op}$ change sign. The lines roughly form a star pattern segmenting the 2D plot into six areas with opposite sign. \textcolor{black}{This six-fold pattern is in agreement with related photoresponse measurements for graphene with a dual-gate layout performed under illumination with near-infrared \cite{Gabor2011,Song2011} and THz light \cite{Castilla2019,Viti2020} and indicates a dominant hot-carrier assisted PTE effect in graphene.} One hence expects that the six-fold segmenting is a property characteristic of PTE. In order to test this supposition, we calculated the contribution of PTE to the responsivity, i.e., the quantity 
$\Delta T_C \cdot (S_{TG2}-S_{TG1})/P_{THz}^{op} $. We assumed a constant carrier temperature difference of $\Delta T_C = 2.8$~K between the TG~1 and TG~2 channel regions (value extracted from \Figrefc{fig:RSMvsPTE}). In \Figrefc{fig:SAC_PTE_MAP}, we present the calculated PTE contribution to the optical responsivity. In this color plot, one also finds the sixfold segmenting of \Figrefb{fig:SAC_PTE_MAP}. 
However, the boundaries between the segments do not quite overlap with the boundaries of \Figrefb{fig:SAC_PTE_MAP}. To clearly show the differences, the dashed lines of \Figrefb{fig:SAC_PTE_MAP} are replotted in \Figrefc{fig:SAC_PTE_MAP}. \textcolor{black}{The observed discrepancy between the calculation and measurement can be attributed to several aspects: First, to the simplified in which we calculated the Seebeck response, without considering the contribution from the ungated region between TG~1 and TG~2. Second, the cancellation of the RSM responses of the two gated regions is not complete for unequal TG~1 and TG~2 voltages and is therefore embeeded in \Figrefb{fig:SAC_PTE_MAP}. Third, we study an antenna-coupled pn-junction, which is not illuminated through a free-space beam and therefore the measured pattern is altered by impedance matching conditions (see Supporting Information), which change for different gate voltages. Finally, such a large-scale responsivity pattern measurement is also expected to be affected by hysteresis \cite{Wang2010}.}

\begin{figure*}[h!t]
\centering
\includegraphics[keepaspectratio, width=1\textwidth]{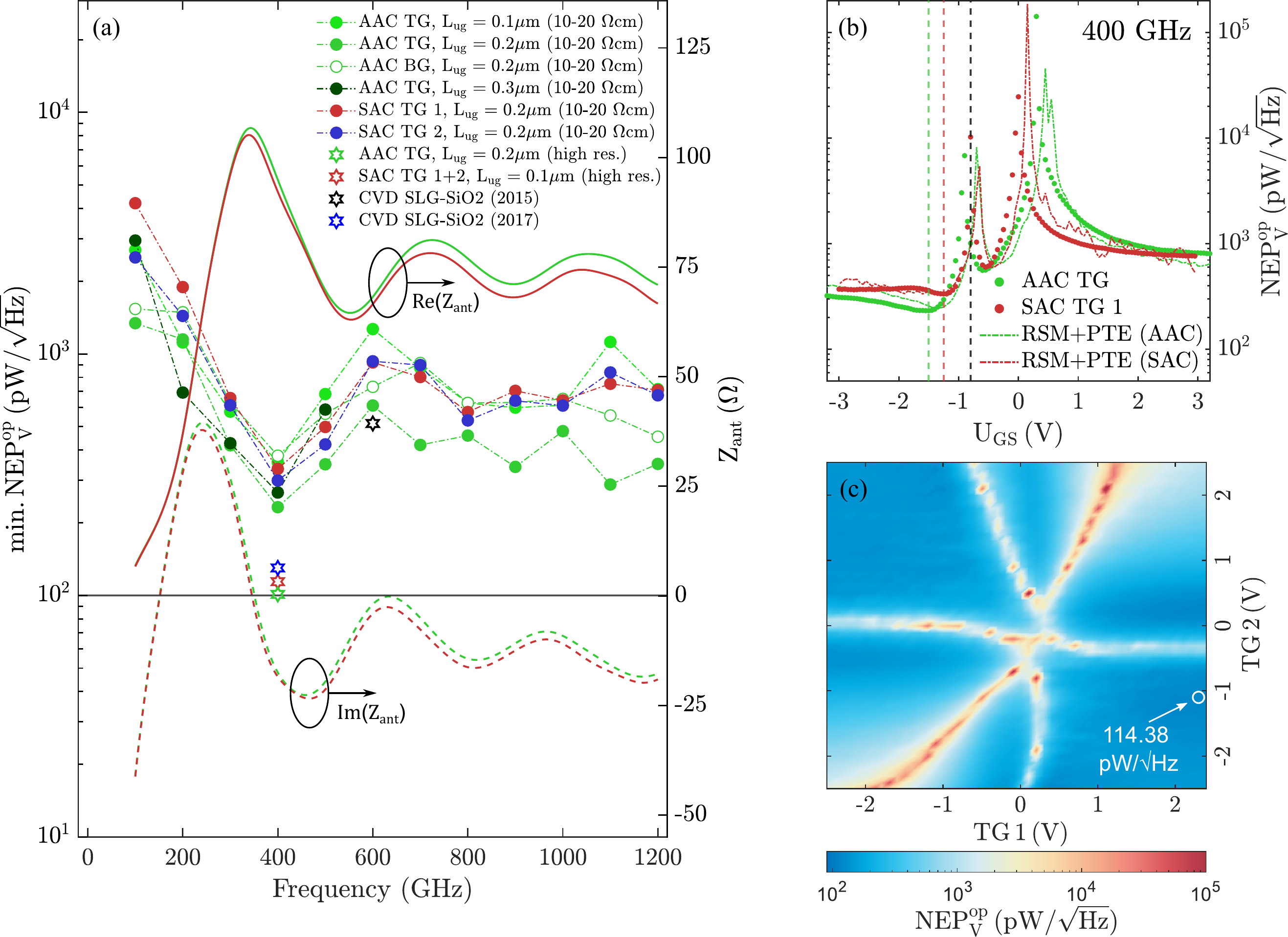}
\caption{(a) Minimum $\rm{NEP}^{op}_V$ (left vertical axis) of multiple G-TeraFETs as a function of frequency for AAC (green dots) and SAC (red for TG~1 and blue for TG~2) in comparison with State-of-the-Art CVD G-TeraFET sensitivity values. Simulated real (solid lines) and imaginary parts (dashed lines) of the antenna impedance $Z_{ant}$ (right vertical axis) for AAC (green) and SAC (red) are also shown. (b) Measured optical noise-equivalent power ${NEP^{op}_V}$ at 400 GHz of a G-TeraFET with asymmetric (AAC, green) and symmetric (SAC, red) antenna coupling. The estimated NEP for AAC and SAC from the linear interpolation model (RSM+PTE, dash-dotted line) are shown in addition. \textcolor{black}{The devices are fabricated on a lightly p-doped substrate with a resistivity of 10-20 $\Omega$cm. $U_{Dirac}$ is marked as vertical black dashed line.} (c) $\rm{NEP}^{op}_V$ (calculated from \Figrefa{fig:SAC_PTE_MAP} and \Figrefb{fig:SAC_PTE_MAP}) at 400~GHz of a G-TeraFET with SAC as a function of the bias voltages applied to TG~1 and TG~2. The best operational point (lowest NEP) is marked with a white circle.}
\label{fig:NEP}
\end{figure*}

\subsection{Noise-equivalent power}
Detector sensitivity can be described by the noise-equivalent-power $\rm{NEP}^{op}_V = V_N/|\Re_V^{op}|$ specifying the input power at which the detector has a unity signal-to-noise ratio limited by (voltage) noise source $V_N$ (in 1 Hz measurement bandwidth). It has been shown that typically the dominant noise source of TeraFETs, because they are operated without drain-to-source bias voltage, is Johnson-Nyquist noise $V_N = \sqrt{4 k_B T R_{DS}}$.\cite{Lisauskas2013,Bauer2014} 

Before comparing the minimal values of $\rm{NEP}^{op}_V$ of various devices in \Figrefa{fig:NEP}, it is important to recall that these minimal values are obtained only at specific optimal bias points. \Figrefb{fig:NEP} shows measured $\rm{NEP}^{op}_V$ curves as a function of the TG voltage for an AAC-type and a SAC-type device at 400~GHz. 
Additionally, curves calculated with the RSM+PTE model are displayed. For both devices, the minimum values of $\rm{NEP}^{op}_V$ are obtained in the hole-conduction branch, where the RSM and PTE signals have the same sign (add constructively). For the AAC-type device, the gate voltage, where we measured the minimal $\rm{NEP}^{op}_V$ value, is $U_{GS} = -1.5~$V (marked as vertical green dashed line), for the SAC-type device, it is $U_{GS} = -1.25$~V (marked as vertical red dashed line), both positions are on the left side from $U_{Dirac} = -0.8~$V. 

For the same two devices -- together with two other AAC devices with different ungated channel lengths $L_{ug} = 0.1$ and $0.3~\mu$m --, we plot in \Figrefa{fig:NEP} the minimal $\rm{NEP}^{op}_V$ values as a function of the radiation frequency. The corresponding gate voltages were found to be only weakly frequency-dependent (see Supporting Information). \textcolor{black}{In the same graph the frequency-dependent real and imaginary parts of the antenna impedance $Z_{ant}$ for our AAC-and SAC-type photodetectors with $L_{ug} = 0.2~\mu$m are presented.} One observes a reciprocal behavior of $Z_{ant}(\omega)$ and the curves of minimal $\rm{NEP}^{op}_V$. The reason is that the performance of the detector strongly depends on the efficiency of the THz power coupling from the antenna to the transistor's graphene channel. Best electrical power coupling is achieved for conjugate impedance matching, $Z_{ant}= Z_{SLG}^{\ast}$, where $Z_{SLG}$ is the gate-voltage-dependent impedance of the graphene channel (the real part of which typically lies in the range of several k$\Omega$, see also Supporting Information). Although we do not reach optimal impedance matching at any frequency, one still clearly sees that the variation of the impedance with frequency directly reflects in the frequency dependence of the minimal $\rm{NEP}^{op}_V$ values.  
At 100~GHz, where $\Re({Z_{ant}})$ is low, the minimal $\rm{NEP}^{op}_V$ is fairly large, with values of 1340~pW/$\surd{\rm{Hz}}$ for the AAC device and 4207~pW/$\surd{\rm{Hz}}$ for the SAC one. 
It is worth mentioning, that for BG voltage operation, the lowest $\rm{NEP}^{op}_V$  (1540~pW/$\surd{\rm{Hz}}$) is close to that obtained for TG biasing. At 1000~GHz, the minimum $\rm{NEP}^{op}_V$ values are significantly smaller 
(479~pW/$\surd{\rm{Hz}}$ for AAC and 642~pW/$\surd{\rm{Hz}}$ for SAC), %.  has decreased by roughly 2.8 for AAC and 6.6 SAC, 
which we can attribute to improved power coupling by a smaller difference in impedance between the antenna and the channel. 
\begin{table*}[!ht]
\footnotesize
\centering
  \caption{Comparison of the room-temperature maximum optical voltage (V) or current responsivity (I) $\Re_{V,I}^{op}$ and minimum optical noise-equivalent power $\rm{NEP}^{op}$ without any normalization (w.n.) at the respective excitation frequencies for different material technologies.}
\begin{tabular}{cccc p{3cm} p{2cm} r} \hline
Material  & Frequency & $\Re_{V,I}^{op}$ & $\rm{NEP}_{V,I}^{op}$ & Antenna & Comment & Ref. \\ 
Technology  & [GHz] & [V/W]  & [pW/$\rm{\sqrt{Hz}}$] &      &         &  \\
& & or [mA/W] & & & & \\
\hline \hline
90-nm Si CMOS & 250 &  408 (V)  & 21  &  \scriptsize  narrowband (AAC) \newline slot & \scriptsize  Si lens (w.n.)  & \onlinecite{Wiecha2021} \\ 

90-nm Si CMOS & 300 &  550 (V)  & 20.8  &  \scriptsize  narrowband (AAC) \newline slot & \scriptsize  Si lens (w.n.)  & \onlinecite{Mateos2020} \\ 

65-nm Si CMOS & 1000 &  765 (V)  & 25  &  \scriptsize  narrowband (AAC) \newline Bi-quad & \scriptsize  Si lens (w.n.)  & \onlinecite{Ferreras2021} \\ 

90-nm Si CMOS & 250-2200 &  220 (V)  & 48/70 \scriptsize (0.6/1.5~THz)  &  \scriptsize  broadband (AAC) \newline log-spiral or bow-tie & \scriptsize  Si lens (w.n.)  & \onlinecite{Ikamas2018} \\ 

\hline \hline
AlGaN/GaN & 504  &  104 (I)  & 25.4 & \scriptsize  broadband (AAC) \newline MIM bow-tie & \scriptsize  Si lens (w.n.)  & \onlinecite{Bauer2019}  \\ 
AlGaN/GaN & 900  &  48 (I)  & 57 & \scriptsize  broadband (AAC) \newline MIM bow-tie & \scriptsize  Si lens (w.n.)   & \onlinecite{Bauer2015}  \\ 
AlGaAs/GaAs & 600  & 55 (I) & 250 &\scriptsize broadband (AAC) \newline logspiral   &\scriptsize Si lens (w.n.)  & \onlinecite{Regensburger2018_CW} \\ 
\hline \hline
CVD Graphene & 400 & 74 (V)  & 130 & \scriptsize broadband (AAC) \newline  split bow-tie  &  \scriptsize  Si lens (w.n.) \newline high resistivity & \onlinecite{Generalov2017}\\ 

CVD Graphene  & 600 & 14 (V) & 515 &  \scriptsize broadband (AAC) \newline split bow-tie &  \scriptsize Si lens (w.n.) \newline high resistivity& \onlinecite{Zak2014}\\ 

single-crystal \\ CVD Graphene &  2800  & 8$^*$ (V) &  600$^*$ &\scriptsize  \scriptsize broadband (AAC) \newline single-gated bow-tie &\scriptsize $^*$normalized to \newline diffraction limited area & \onlinecite{Asgari2021} \\ 

hBn-encapsulated \\ CVD Graphene &  2800  & 4.5$^*$ (V) &  2000$^*$ &\scriptsize  \scriptsize broadband (AAC) \newline single-gated bow-tie &\scriptsize $^*$normalized to \newline diffraction limited area & \onlinecite{Asgari2022} \\ 

CVD Graphene  & 400 & 36 (V) & 232 &  \scriptsize broadband (AAC) \newline split bow-tie &  \scriptsize Si lens (w.n.) \newline 10-20 $\Omega$cm&  \textbf{This work}\\ 

CVD Graphene  & 400 & 30 (V) & 299 &  \scriptsize broadband (SAC) \newline double split \newline bow-tie &  \scriptsize Si lens (w.n.) \newline 10-20 $\Omega$cm&  \textbf{This work}\\ 

CVD Graphene  & 400 & 59 (V) & 101 &  \scriptsize broadband (AAC) \newline split bow-tie &  \scriptsize Si lens (w.n.) \newline high resistivity&  \textbf{This work}\\ 

CVD Graphene  & 400 & 63 (V) & 114 &  \scriptsize broadband (SAC) \newline double split \newline bow-tie &  \scriptsize Si lens (w.n.) \newline high resistivity &  \textbf{This work}\\ 

\hline
exf. SLG-hBN &  2520  & 105$^*$ (V) &  80$^*$ &\scriptsize broadband (SAC) \newline broad-wire dipole    &\scriptsize $^*$normalized to \newline diffraction limited area  & \onlinecite{Castilla2019} \\

exf. SLG-hBN &  2800  & 49$^*$ (V) &  160$^*$ &\scriptsize  \scriptsize broadband (AAC) \newline split bow-tie   &\scriptsize $^*$normalized to \newline diffraction limited area  & \onlinecite{Viti2020} \\ 

exf. SLG-hBN &  3400  & 50$^*$ (V) &  120$^*$ &\scriptsize  \scriptsize broadband (SAC) \newline broad-wire dipole   &\scriptsize $^*$normalized to \newline diffraction limited area  & \onlinecite{Viti2021} \\

exf. SLG-hBN &  300  & 1.9$^*$ (I) &  670$^*$ &\scriptsize  \scriptsize asymmetric double grating-gate periodic nanostructure \newline   &\scriptsize $^*$normalized to \newline diffraction limited area  & \onlinecite{Delgado2022asymmetric} \\

\hline\end{tabular}\label{tab:NEPvalues}%y
\end{table*}%
Clearly, the lowest $\rm{NEP}^{op}_V$ values are obtained at 400~GHz, close to the maximum magnitude of the antenna impedances, where the best power coupling is achieved. \textcolor{black}{It is worth noting that close to this frequency, also the field enhancement factor $E_{||}$ ($\sim$1500 at 400 GHz) of the two antenna layouts peaks (see Supporting Information, Fig. S6(b)). Larger values of $E_{||}$ lead to an increased carrier heating in the antenna gap, which is advantageous for the detector performance.} For the AAC-type device, we obtain a minimal $\rm{NEP}^{op}_V$ of 232~pW/$\surd{\rm{Hz}}$, 80\% larger than the record value of 130~pW/$\surd{\rm{Hz}}$ achieved with a CVD G-TeraFET, also at 400~GHz (data point shown as blue open star in \Figrefa{fig:NEP})\cite{Generalov2017}. The latter value was obtained with a device fabricated on an undoped (high-resistivity) substrate. As shown in the Supporting Information (Fig. S2), the substrate doping leads to Drude absorption of the THz radiation, attenuating the beam by a factor of 0.66 (determined experimentally via transmission measurements). 
Multiplying the measured $\rm{NEP}^{op}_V$ value of 232~pW/$\surd{\rm{Hz}}$ with the attenuation factor yields a hypothetical $\rm{NEP}^{op}_V$ value of $0.66 \cdot 232$~pW/$\surd{\rm{Hz}} = 153$~pW/$\surd{\rm{Hz}}$ for an absorption-free Si substrate, close to the record value from above. 
Finally, we present two more data points in \Figrefa{fig:NEP}, which we obtained with devices on undoped substrate upon further optimization of the contact resistance. 
We achieved a minimum $\rm{NEP}^{op}_V$  of 101~pW/$\surd{\rm{Hz}}$ with an AAC-type device (open green star) and of 114~pW/$\surd{\rm{Hz}}$ with an SAC-type one (open red star), both at 400~GHz. For the latter, the best operational point was extracted from the 2D TG~1/TG~2-biasing map of $\rm{NEP}^{op}_V$ presented in \Figrefc{fig:NEP}. 
The reduced contact resistance of the devices led to a peak $R_{DS}(U_{GS})$ resistance of 5.5~k$\Omega$ rather than the 7~k$\Omega$ of non-optimized devices. To the best of our knowledge, both $\rm{NEP}^{op}_V$ values represent the lowest values reported for CVD G-TeraFETs to date. 

\section{Discussion}
To put our results more into perspective with the state-of-the-art of THz photodetectors regarding their NEP values, we provide in Table~\ref{tab:NEPvalues} an overview of important results reported about in the literature. The list focusses on graphene-based devices, summarizing the leading work of which we are aware. The devices were fabricated from either CVD graphene or exfoliated graphene (the notation ``exf. SLG-hBN'' in the table specifies that the SLG is encapsulated in hBN). Not only detectors for THz frequencies are listed but also devices which have been optimized as thermoelectric detectors for infrared radiation. In addition, we present a selected number of results achieved with Si CMOS and GaN/AlGaN TeraFETs.

In the column entitled ``Comment'', we distinguish between two types of measurement and evaluation approaches. The comment ``Si lens (w.n.)'' stands for the determination of true \textit{optical} responsivity and NEP values, where the power $P_{THz}^{op}$ in the expression of the responsivity is the total as-measured beam power. The radiation was coupled onto the detector through a Si substrate lens. No further data processing was made (``w.n.'' standing for ``without normalization''). On the other hand, ``normalized to diffraction-limited area'' stands for the determination of a \textit{cross-sectional} responsivity, respectively NEP value (see footnote 1). Here, only a part of the beam power contributes to the measured rectified signal (for example because no substrate lens could be used). Various approaches are used to determine this part by experiment or simulations. Often it is assumed that the beam which has been focused by a lens or mirror, has a diffraction-limited size at the focal position where the detector is placed. Based on this assumption, the NEP value is calculated which one would obtain if the full beam power would contribute to the rectification\cite{Javadi2021,Bauer2019}. In Table~\ref{tab:NEPvalues}, we mark cross-sectional responsivity and NEP values with an asterisk.

The central information of Table~\ref{tab:NEPvalues} is that graphene THz detectors with broadband antenna structures are now reaching NEP values on the order of 100~pW/$\surd{\rm{Hz}}$ at sub-1-THz frequencies, where they enter a regime which previously was only occupied by detectors based on classical semiconductors (notably Si CMOS and AlGaN/GaN). However, they do not yet surpass the performance level of these, which reach NEP values of below 50~pW/$\surd{\rm{Hz}}$ as broadband devices. The on-going progress in graphene device technology will most likely lead G-TeraFET performance gradually catching that of classical semiconductor TeraFETs. For example, advances are to be expected with regard to the continued reduction of contact resistances and by improved detailed device design leading to better impedance matching. But progress may also come in less obvious ways. A possible route where such progress may occur relates to electron-phonon interaction, specifically acoustic phonon scattering of the charge carriers. One aspect which has an impact on PTE. As already mentioned above, the scattering rate for exfoliated graphene is usually low (the momentum relaxation time long), ensuring a long cooling time of the charge carriers, which is beneficial for PTE. \textcolor{black}{Our EM wave simulations (see Supporting Information Fig. S4) indicate that higher carrier mobility is likely to increase the amount of dissipated THz power in the gated channel region. However, to investigate the change in the final RSM-to-PTE ratio for an increased carrier mobility, the interplay between (i) the increased electronic cooling length \cite{Antidormi2021,Vangelidis2022}, (ii) the increased Seebeck coefficient \cite{Duan2016,Miseikis2020}, and (iii) the change in the power dissipation profile (see \Figrefc{fig:RSMvsPTE} and Fig. S4(a) in the Supporting Information) need to be taken into account. The actual RSM-to-PTE ratio in high mobility G-TeraFETs can be studied using similar methods as presented in this work, where the detectors are fabricated with e.g. hBN encapsulated graphene. 
While there is still more room to investigate the RSM-to-PTE ratio in high mobility graphene transistors, based on the computational and experimental results of this work, we can conclude that for CVD-grown graphene THz detectors with low mobility and SAC- and AAC-type detector layouts, the PTE is expected to be the dominant rectification mechanism.}

 \section{Conclusion}
\textcolor{black}{In summary, it has been demonstrated that the detector response of G-TeraFET is a combination of resistive self-mixing (RSM) and thermoelectric (PTE) rectification. We have also shown that in case of G-TeraFETs fabricated with low-mobility CVD-grown graphene the RSM-to-PTE ratio varies strongly with the frequency of the incident THz-field and that the PTE is the dominant rectification mechanism at higher frequencies. For our CVD G-TeraFETs with asymmetric antenna coupling, the PTE already dominates over the RSM above 100 GHz. With EM simulations we showed that the observed frequency dependence is a direct consequence of the relative change in the total dissipated THz power (the surface loss density) between the gated and ungated graphene channel regions. From the simulations we also found that the channel length over which the PTE contributes to the photoresponse below the gate electrode corresponds approximately to the electronic cooling length. In addition, a second, weaker PTE was identified that can be attributed to the contact doping effect in graphene near the metal electrodes. At 400 GHz our detectors achieve a minimum optical noise-equivalent power of 114 pW/$\sqrt{\rm{Hz}}$ for symmetric and 101 pW/$\sqrt{\rm{Hz}}$ for asymmetric antenna coupling. This work demonstrates the potential of harnessing the PTE response in G-TeraFETs and paves the way for the next generation high performance graphene THz detectors.}
\newpage
\section*{Methods}
\subsection*{Device fabrication}
The CVD G-TeraFETs were fabricated either on 6" $p$-doped Si wafers (\textcolor{black}{350 \si{\um} thick,} 10-20~$\Omega$cm) with 90-nm-thick, thermally grown SiO$_2$ layer, or on 6" highly resistive Si wafers (\textcolor{black}{350 \si{\um} thick,} >10~k$\Omega$cm) with a 300-nm-thick thermally grown SiO$_2$ layer. %\JH{Isn't only the doped version covered with the 90 nm substrate, or was that the case for both? AG: Yes, highresistive wafers have 300 nm SiO_2}. 
The wafers were patterned with lithography markers and contact pads and the commercially available CVD graphene from Graphenea was transferred onto the substrate. To encapsulate the graphene for protection against contamination by the liquid chemicals used throughout the rest of the fabrication process, a seed layer of Al was deposited onto the graphene surface and oxidized to Al$_2$O$_3$. Next, the Ohmic source and the drain contacts were patterned via electron-beam lithography, the protective dielectric at the contact areas was etched, and \textcolor{black}{the top contacts were evaporated with Cr 5 nm/Au 100 nm stack. The evaporation chamber was pumped overnight to reach pressure below $10^{-7}$ bar and the evaporation rate of 1 Å/s was used. Next, the devices were covered by a 18-nm-thick ALD-grown Al$_2$O$_3$ gate dielectric, the TG was patterned via electron beam lithography and Ti 5nm /Au 100 nm stack was evaporated. The whole process was performed on full wafers, as shown in \Figrefe{fig:schematics+IV}.} 

About 1000 GFETs were fabricated in each batch, tested on-wafer using an automated DC probe station, then the wafer was diced, about 30 chips were packaged, transported, and bonded on a PCB readout board for measurements.
\textcolor{black}{A statistical analysis of the performance variations of the detectors \cite{Boppel2011} is a task of future studies. Here, we refrain from this analysis for the following reason. While our wafer-scale fabrication of the GFETs comes with a moderate standard deviation of the detector performance, and a similar spread of the contact resistances and mobilities, we find it difficult to identify a dependency of the detector performance on fabrication parameters. An attempt at a causality analysis will be confusing and inconclusive based on our limited data presently available. For the sake of completeness, we provide the following information: We measured the DC electrical transport characteristics of $N=57$ devices on the same wafer of the devices presented in the main text, and obtained (using the fitting model presented in the next paragraph) a mean electron mobility of $\bar{\mu}_{FE,e} = 1924$~cm$^2$/Vs with a standard deviation $\sigma_s({\mu}_{FE,e})=523$~cm$^2$/Vs, and correspondingly for holes $\bar{\mu}_{FE,h} = 1414$~cm$^2$/Vs with $\sigma_s({\mu}_{FE,h})=390$~cm$^2$/Vs. }
\subsection*{Fitting model}
In this work, we adapted the fit procedure presented in Kim et al. \onlinecite{Kim2009} for our multi-gate transistors. 
The fit model uses a semi-empirical square-root dependence of the carrier concentration on the gate voltage (ether TG or BG) and accounts for the quantum capacitance of the 2D electrons in graphene. The total resistance of the channel is fitted continuously through the Dirac cone - when switching from electron ($\alpha=1$) to hole conduction branch ($\alpha=-1$) - and in addition we extract total contact resistance $R_C$ of the devices directly via 
\begin{equation}
\begin{split}
    R_{DS}(U_{GS}) & = \frac{R_C \cdot N_g \frac{L_g}{L_{ug}} + N_{ug} \cdot R_{DS}(0 \,\mathrm{V})}{N_{ug} + N_g\frac{L_g}{L_{ug}}} \\
    & + \frac{N_g}{e(n_{2D,e} \,\mu_{FE,e}+n_{2D,h} \, \mu_{FE,h})} \,\frac{L_g}{W} \,,
\label{eq:Rfit_model}
\end{split}
\end{equation}
where $n_{2D,\alpha}=({n_0}^2+[n_\alpha(U_{GS})]^2)^{1/2}$ and $R_{DS}(0 \,\mathrm{V})$ is the measured resistance at zero gate bias. The gate voltage dependent carrier density in the graphene layer $n_\alpha(U_{GS})$ can be estimated from
\begin{equation}
\begin{split}
    n(U_{GS}) & = \bigg(n_{q}^2 + \bigg[\frac{C_{ox,TG}}{e}(U_{GS}-U_{Dirac}) \\ & \cdot \Theta(\alpha(U_{GS}-U_{Dirac}))\bigg]^{2}\bigg)^{-1/2} - n_q
    \label{eq:nfit_model}
\end{split}
\end{equation}
where $\Theta(...)$ denotes the Heaviside function and $n_{q}=\pi(\hbar v_F C_{ox,TG})^2/(2e^4) $ accounts for the quantum capacitance effect. $U_{Dirac}$ is set as a fixed parameter here and determined from the gate voltage point of maximum resistance. Note that the above model can be employed for BG operation if $C_{ox,TG}$ is replaced by the BG capacitance $C_{ox,BG}$. 
\subsection*{Simulations of power dissipation}
We provide here details of the EM wave simulations (with CST) of the THz power dissipation in the graphene channel of AAC-type devices. The surface conductivity/surface impedance of the graphene was calculated by means of the Kubo formalism \cite{Falkovsky2007,Llatser2012}. We assumed a finite SLG thickness of $\Delta_{SLG} = 10$~nm, a constant chemical potential of $\mu = 0.2$~eV and a momentum-relaxation time of $\tau_p = 60$~fs for CVD-grown graphene\cite{Soltani2020}. Note that further simulations, where $\tau_p$ is varied from 60 over 180 to 300 fs, are presented in the Supporting Information (Fig. S4).  The antenna was excited by a plane-wave through the silicon substrate to mimic the backside illumination through a silicon lens in our experiments. 
The graphene sheet is oriented in $x$ ($\in$ [0,L]), $y$ ($\in$ [0,W]) and $z$ ($\in$ [$-\Delta_{SLG}$,0]) (as illustrated by the red coordinate frame in the inset of \Figrefc{fig:RSMvsPTE}). The surface loss density SLD is given by the power flow per unit area of a plane wave flowing towards the graphene sheet (negative z direction) via\cite{Orfanidis2016} 

\begin{equation}
   P_D(x,y,0,\omega) = \frac{1}{2} R_s |\textbf{H}(x,y,0,\omega)|^2,
\end{equation}
where, $R_s$ denotes the surface resistance of the graphene sheet determined from the real part of surface impedance and $\textbf{H}(x,y,0,\omega)$ is the in-plane magnetic field amplitude at the position z=0 of SLG surface. In order to estimate the relative amount of power dissipated in the gated and ungated channel regions we use the SLD extracted at SLG channel center (y=W/2,$P_D(x,W/2,0,\omega)=P_D(x,\omega)$). 

\subsection*{Measurement setup}
\begin{figure}[h!t]
\centering
\includegraphics[keepaspectratio, width=0.45\textwidth]{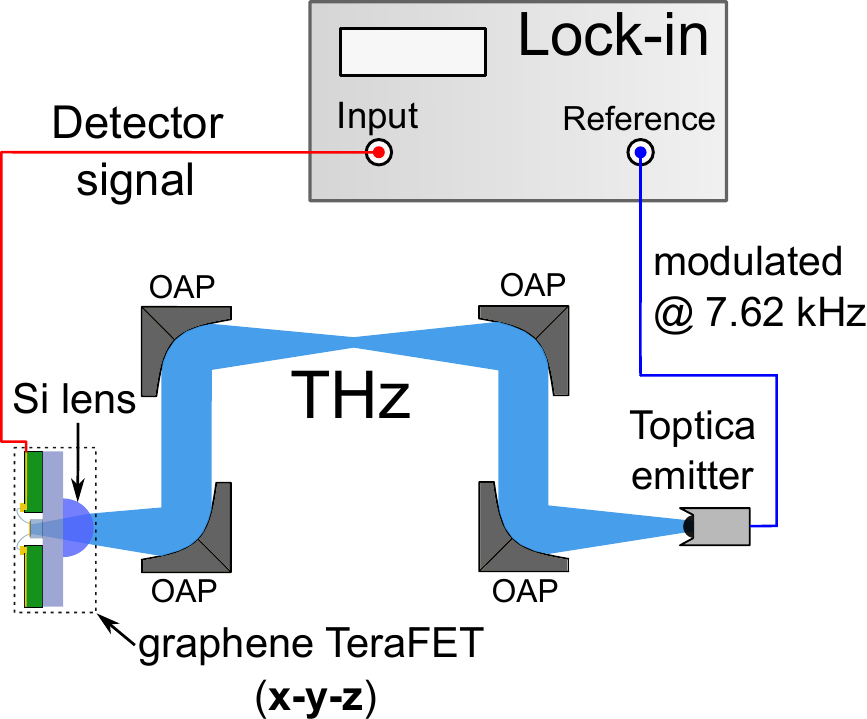}
\caption{Schematic of the measurement setup for the THz characterization of the graphene TeraFETs. The THz beam is guided through four off-axis parabolic mirrors towards the detector.}
\label{fig:Setup}
\end{figure}

The DC electrical characterization was performed using the Keysight B2900 precision source/\-measure unit. The drain-to-source voltage $U_{DS}$ needed for the read-out of the drain-to-source current $I_{DS}$ was fixed at 10~mV, S electrode was grounded. 
For the determination of the THz responsivity $\Re_V^{op}$ of the TeraFETs, we employed a commercial CW spectroscopy system (Toptica TeraScan 1550) which has a tunable fiber-coupled InGaAs photomixer as the radiation source, and replaced the photomixer detector of that system by our TeraFET. \textcolor{black}{All measurements were performed at room temperature and under ambient conditions. As shown in \Figref{fig:Setup} the \textcolor{black}{TeraFETs} were illuminated through the substrate to which a hyperhemispherical Si lens (diameter: 12~mm, height: 7.1~mm) was attached (not glued). For THz alignment the TeraFETs were mounted on a \textit{x-y} micrometer translation stage, while the lens was fixed in its position. The Si lens $+$ G-TeraFETs were mounted on an additional \textit{z}-stage to adjust the distance from the last off-axis parabolic mirror (a 3" OAP) and to guarantee that the devices were characterized in the focal point of the OAP. At this position the free-space power of the THz beam ($P^{op}_{THz}$) was determined with a calibrated large-area Golay cell \textcolor{black}{(SN 160735, Microtech Instruments)}.} The rectified voltage ($\Delta U_{DS}$) was measured using the lock-in technique (DSP Model 7265) at a modulation frequency of 7.62~kHz. For that purpose, the THz radiation was chopped electronically by a bipolar sinusoidal modulation of the electrical bias on the InGaAs photomixer gap of the Terascan 1550 system. The peak-to-peak photoresponse $\Delta U_{DS}$ for our measurement system has been obtained from $\Delta U_{DS} \approx 3 \cdot U_{lock-in}$\cite{Ferreras2021}.

\section*{Supporting Information}
\textcolor{black}{Supporting Information. Contents: PTE power dependency, absorption losses in the lightly p-doped Si substrate, frequency dependence of the
gate voltage of minimal NEP, back gate responsivity measurements, carrier mobility dependence on the RSM-to-PTE responsivity ratio, frequency dependence of
the extracted RSM and PTE responsivity, an extensive antenna analysis including the field enhancement factor, the effect of back gate voltage on the AAC detector
performance and estimation of the prefactors for RSM and PTE.}

\bibliographystyle{unsrt}  
\bibliography{THz}

\section*{Acknowledgements}

We acknowledge financial support by DFG grants RO 770/40-1 and -2 (Germany), and the Academy of Finland through grants no. 314809 (LAMARS), 343842 (Postdoctoral researcher A. Generalov), and 342586 (HyPhEN) and QTF Centre of Excellence project No. 336817.

\section*{Author contributions statement}

H.G.R., A.G. and F.L. conceived the experiment, F.L., A.G., J.H. and M.P. contributed to detector design, A.G., A.M. and K.V. fabricated the devices, F.L. and J.H. conducted the THz measurements and F.L. analyzed the results. F.L., A.G., M.P. and H.G.R. wrote the manuscript which was reviewed by all authors. 

\section*{Competing interests}
The authors declare no competing financial interest.

\newpage
\title{Supporting Information \\ Terahertz detection with graphene FETs: photothermoelectric and resistive self-mixing contributions to the detector response}

%\section{\textit{Supporting Information}  \\-- Terahertz detection with graphene FETs: photothermoelectric and resistive self-mixing contributions to the detector response}
\maketitle
\renewcommand{\thepage}{S\arabic{page}} 
\renewcommand{\theequation}{S\arabic{equation}} 
\renewcommand{\thesection}{S\arabic{section}}  
\renewcommand{\thetable}{S\arabic{table}}  
\renewcommand{\thefigure}{S\arabic{figure}}

\addcontentsline{toc}{chapter}{Supporting Information}

\setcounter{page}{1}

\setcounter{section}{0}
\renewcommand*{\theHsection}{chX.\the\value{section}}
\setcounter{figure}{0}
\setcounter{equation}{0}

\section{Power dependency of the PTE}

\begin{figure*}[h!t]
\centering
\includegraphics[width=5.2in]{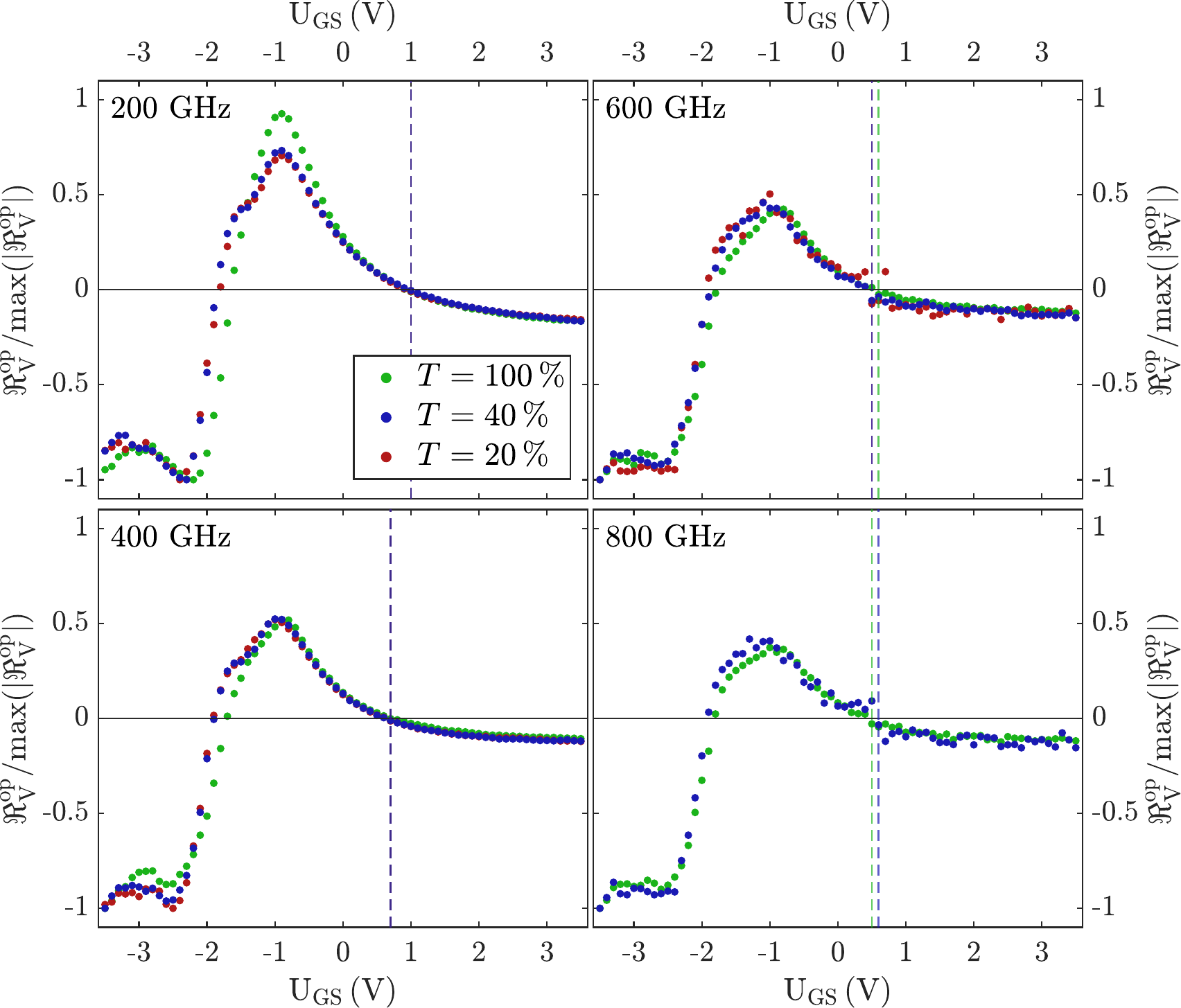}
\caption{THz power dependence (as function of different filter transmission) of the normalized voltage responsivity for an AAC device with $L_{ug} = 0.1 \mu$m at 200, 400, 600 and 800 GHz. The measurements indicate no power dependence on the position of the second phase change for which the PTE is responsible.}
\label{fig:supple_power_dep}
\end{figure*}

In \Figrefc{fig:supple_power_dep}, we present the THz power dependency of a similar AAC device with $L_{ug} = 0.1 \mu$m at different THz frequencies using three different THz filters with transmission factors (T) of 20\%, 40\%, 100\% in order to verify that the observed strong frequency-dependency of the PTE in this work in not is not related to the exponential power decay of our Toptica TeraScan 1550 system as presented in Fig.~3(b). We observe no dependency on the position of the second phase change on the electron conduction branch - which can be related to the PTE for an n-doped CVD graphene TeraFET - at the seperate excitation frequencies (200, 400, 600 and 800 GHz) when changing the relative optical THz input power from 100\% down to 20\%. Therefore we excluded that the PTE frequency-dependence - reported in this work - is related to our Toptica TeraScan 1550 output power.

\section{Absorption losses in the lightly p-doped Silicon substrate}

\begin{figure*}[h!t]
\centering
\includegraphics[width=4.3in]{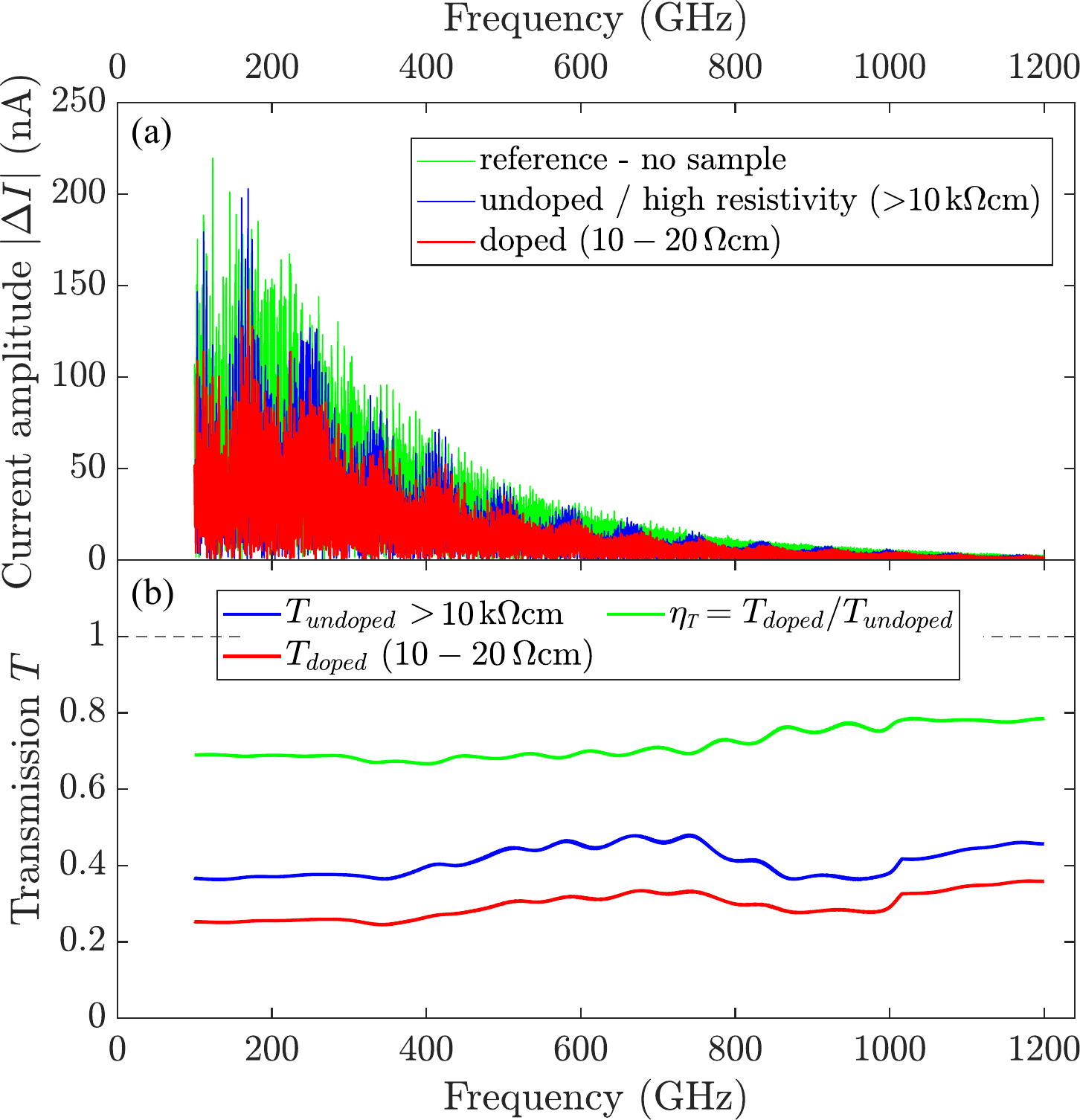}
\caption{(a) Measured current amplitude for no sample (reference, green line), lightly p-doped silicon (10-20$\Omega$cm, red line) and high-resistivity (>10k$\Omega$cm, blue line) silicon substrate. (b) Extracted transmission coefficients $T$ after Fourier filtering of the signal presented in (a) for both type of wafers and calculated additional losses $\eta_{T}$ in the lightly p-doped substrate, which has been used in this work.}
\label{fig:supple_absorption}
\end{figure*}

In \Figref{fig:supple_absorption}, we present measurements of the THz transmission through two silicon substrate with 525$\mu$m thickness. All measurements were performed with an InGaAs photomixer receiver from Toptica (Toptica TeraScan 1550). Measurements were performed in nitrogen atmosphere to reduce the effects of water absorption lines in our transmission analysis. Note that, for the determination of the transmission, the silicon substrates were placed in an intermediate focal point and the receiver was placed in the final focal of our THz characterization system point (as the CVD graphene TeraFET detectors). The Transmission coefficients - presented in \Figrefb{fig:supple_absorption} - were determined after Fourier filtering of the raw data shown in \Figrefa{fig:supple_absorption}. Finally, we determine the losses due to the present doping in our silicon substrates from the ratio of both extracted transmission factors $\eta_{T} = T_{doped}/ T_{undoped}$.

\section{Frequency dependence of the gate voltage of minimal NEP}
The gate voltage $U_{GS,min}$ where one finds the minimal value of $\rm{NEP}^{op}_V$, exhibits a weak dependence on the radiation frequency, which is a consequence of the frequency dependence of the impedances of the antenna and the graphene channel. The minimal value lies always on the hole-conduction side of the gate voltage. Here's some examples for TG-biased AAC-type detectors, fabricated on lightly $p$-doped substrates (10-20 $\rm{\Omega}$cm). For a device with $L_{ug} = 0.1~\mu$m, the minimal $\rm{NEP}^{op}_V$ is found at $U_{GS,min} = -0.9$~V at 100~GHz and at $-1.2$~V at 900~GHz. For a device with $L_{ug} = 0.2~\mu$m, correspondingly at $U_{GS,min} =-1.45$~V at 100~GHz and at $-1.55$~V at 1000~GHz. For a device with $L_{ug} = 0.3~\mu$m, $U_{GS,min} = -1.25$~V at 100~GHz and $-1.35$~V at 400~GHz. 

\section{Back gate responsivity measurements}
\begin{figure*}[h!t]
\centering
\includegraphics[width=5.2in]{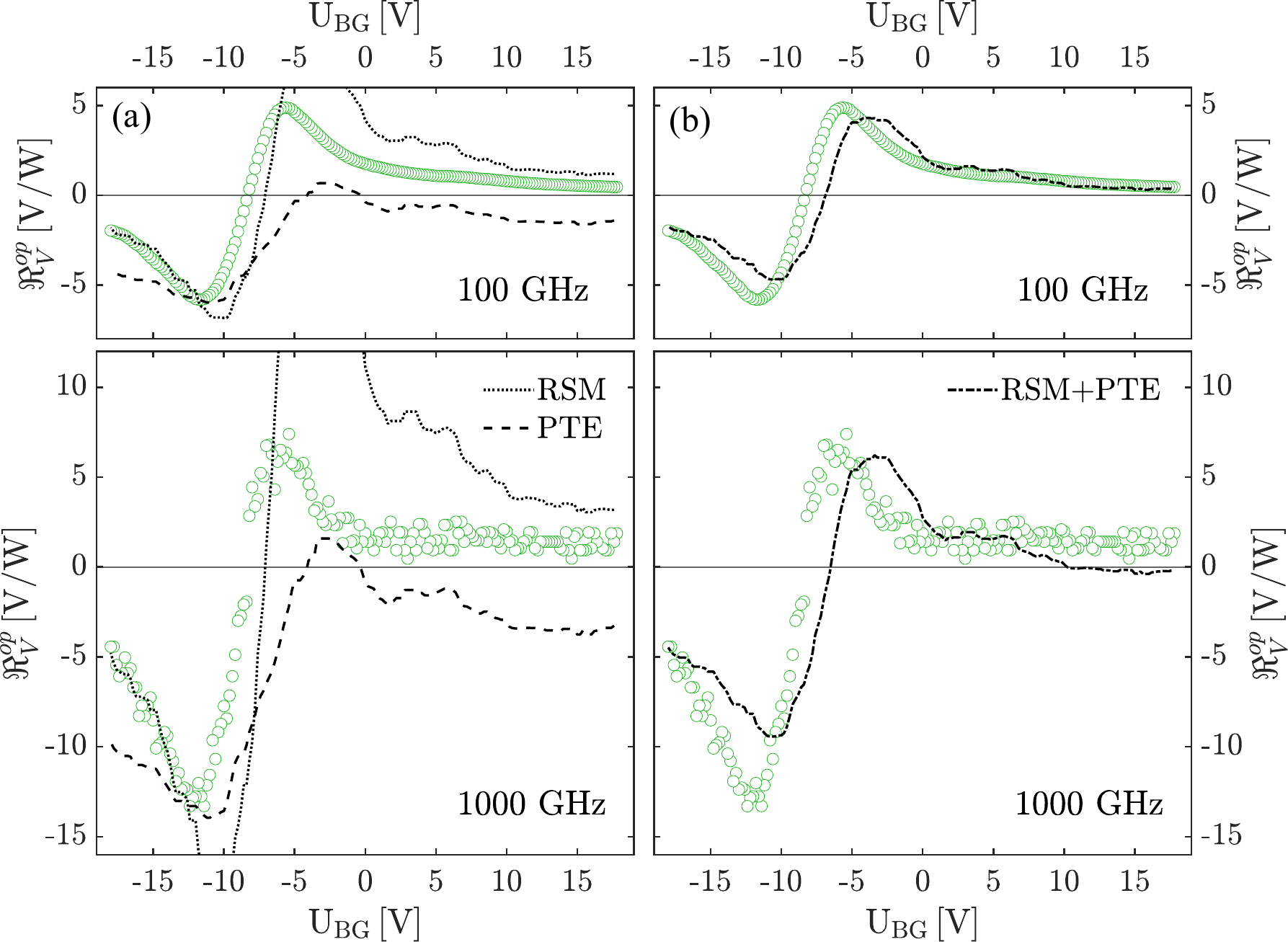}
\caption{(a) Measured optical voltage responsivity ${\Re_V^{op}}$ of CVD graphene TeraFETs with asymmetric antenna coupling  (AAC) and $L_{ug} = 0.2~\mu$m (green open circles) as a function of BG voltage. The calculated photoresponse for resistive self-mixing (RSM, dotted lines, calculated with Eqn.~(5)) and photothermoelectric rectification (PTE, dashed lines, calculated with Eqn.~(1) together with Eqn.~(2)) are shown in addition. (b) Measured and calculated optical responsivities. The calculated curves were obtained by fitting a linear combination of the theoretical RSM and PTE photoresponses (RSM+PTE, dash-dotted lines) to the measured curves.}
\label{fig:supple_BG_measurement}
\end{figure*}

In \Figrefa{fig:supple_BG_measurement} the measured optical THz responsivities at 100 (top) and 1000 GHz (bottom panel) as a function of BG voltage for an AAC type graphene TeraFET with $L_{ug} = 0.2~\mu$m (green open circles) are depicted. We test the measured photoresponse against the theoretical predictions based on the RSM mechanism and the PTE mechanism. Both calculated from the measured BG conductivity $\sigma (U_{BG})$. As shown in \Figrefa{fig:supple_BG_measurement}, nether the RSM nor the PTE can quantitatively explain the measured detector responsivity at the different frequencies. For simplicity, we assumed that the graphene channel in the contact region is Fermi level pinned by the metal electrode, such that $S_g(U_{BG} = 0 \rm{V}) = S_{ug}$. In this case the PTE photoresponse is given by $U_{PTE}^{BG} \approx \Delta T_C \cdot(S_{ug}^{hot} - S_{g}(U_{BG}))$. Following the considerations in \cite{Chaves2015}, a more appropriate way to model the graphene channel in contact with the Source metal electrode under BG voltage tuning would be to treat it as partially (indicated by the ${'}$ symbol) BG voltage dependent ($S_{ug}(U_{BG}^{'})$). In their work the authors modeled the dipole layer between graphene the contact metal as a very thin insulator with a metal-to-SLG equilibrium distance in the orders of 2-4 \AA. Therefore the latter functionality for $E_F(U_{BG}^{'})$ implies exact knowledge of several electrical parameters of the Source metal, e.g. the metal workfunction, the metal-to-SLG equilibrium distance as well as the potential drop in the dipole layer between SLG and the metal.  \\
In \Figrefb{fig:supple_BG_measurement} the data is compared to a linear combination of the theoretical RSM and PTE photoresponses (RSM+PTE). We find semi-quantitative agreement between the linear combination model the measured photoresponse, which shows that also in BG voltage operation the detector signal can be broken down to a combination of the RSM and the PTE. However, when compared to the TG measurements in Fig.~2(b), the overall quality of the fit is reduced for the BG case, which we attribute to our simplified assumptions on $S_{ug}$ for the PTE calculation (see above). At 1000 GHz, the linear combination model predicts a second sign change of the photoresponse for larger positive BG voltages (as one finds for the TG measurements, see Fig.~2(b)). This is not observed for the BG measurements. We speculate that the Seebeck difference between the graphene channel in the contact region and that of the "fully" back gated graphene channel further away from the source electrode is reduced ($S_{ug}(U_{BG}^{'}) \approx S_{g}(U_{BG}))$) when larger positive (and negative) BG voltages are approached. This could explain why the RSM photoresponse dominates over the PTE in these BG voltage regimes.

\section{Carrier mobility dependence on the RSM-to-PTE responsivity ratio}
\begin{figure*}[h!t]
\centering
\includegraphics[width=5.2in]{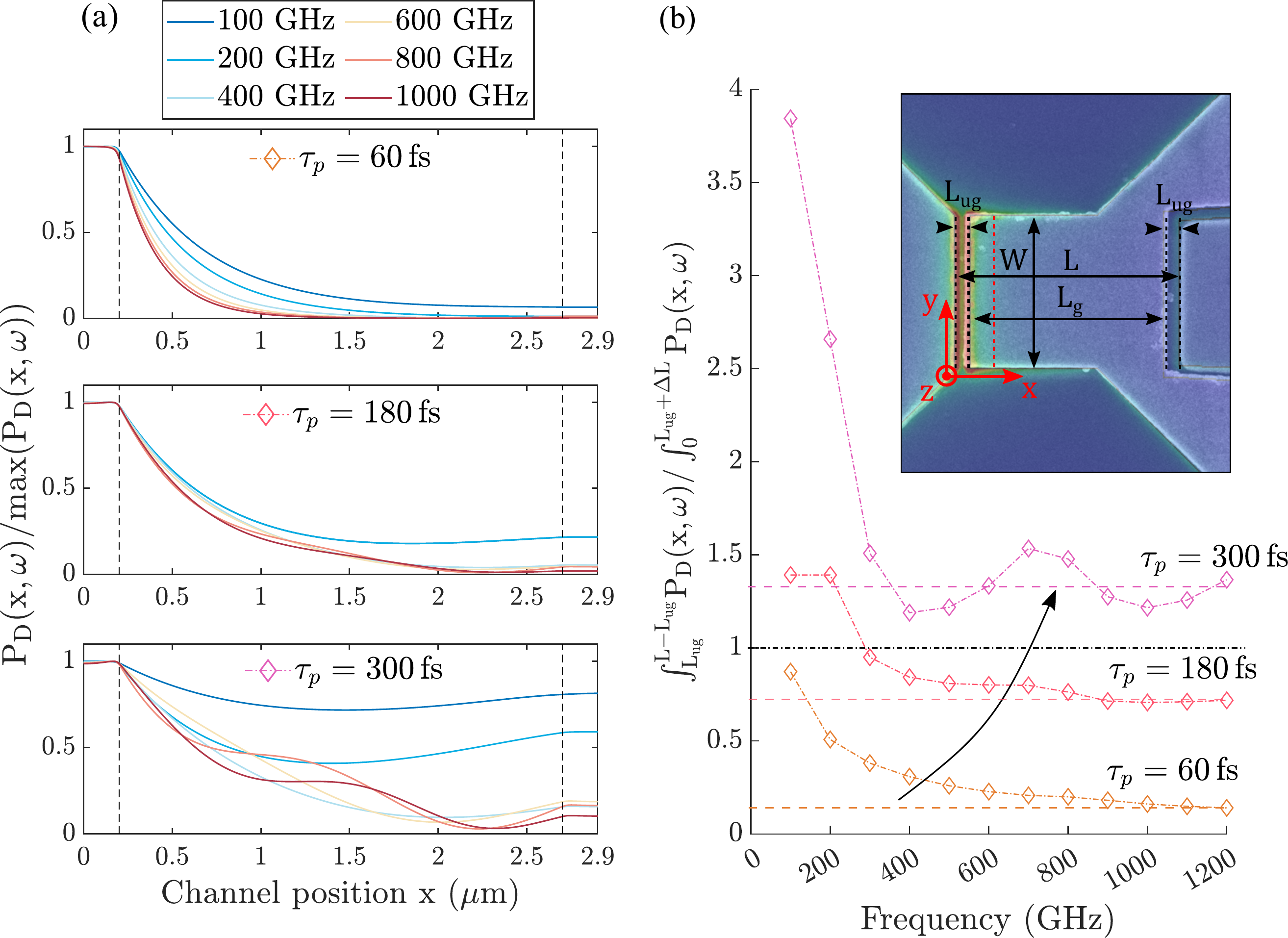}
\caption{(a) Calculated normalized surface power loss density $P_D (x,\omega)$ along the graphene channel ($x$-direction) as a function of THz frequency for a AAC-type device with $L_g=L_{g,AAC}=2.5~\mu$m and $L_{ug}=0.2~\mu$m assuming different momentum relaxation times $\tau_p$ (60fs, 180fs and 300fs). The simulations were performed with CST Studio Suite. 
(b) CST-simulated ratio of the total surface power loss density attributed to RSM vs. that attributed to PTE. The inset depicts a SEM picture of the AAC detector's channel region overlayed by the simulated field distribution at 400~GHz and the coordinate system (shown in red).}
\label{fig:supple_mobility_dependence}
\end{figure*}

\Figrefa{fig:supple_mobility_dependence} depicts the calculated normalized surface power loss density (SLD) $P_D (x,\omega)$ along AAC-type device with $L_g=L_{g,AAC}=2.5~\mu$m and $L_{ug}=0.2~\mu$m assuming three different momentum-relaxation times $\tau_p$ (60, 180, 300 fs) to calculate the surface conductivity/surface impedance of the graphene sheet based on the Kubo formalism \cite{Falkovsky2007,Llatser2012}. For all simulations we assumed a finite SLG thickness of $\Delta_{SLG} = 10$~nm, a constant chemical potential of $\mu = 0.2$~eV. From \Figrefa{fig:supple_mobility_dependence} it is clear that the higher the mobility (corresponding to larger values of $\tau_p$) of the graphene sheet, the more THz power is dissipated along the gated channel region. From the simulated surface power loss density $P_D (x,\omega)$ we calculate the SLD-integral ratio for the transistor part, which attributes to the PTE recitification (from $x=0$ to $x=L_{ug} + \Delta L$) and that which contributes to the RSM signal (from $x=L_{ug}$ to $x=L_{ug} + L_{g}$). For all calculations the effective PTE penetration \textcolor{black}{length} $\Delta L$ underneath the gate electrode was assumed to be $0.3~\mu$m. \textcolor{black}{It should be stressed here that $\Delta L$ is expected to increase sublinearly with increasing carrier mobility as it is linked to the electronic cooling length of the charge carriers \cite{Antidormi2021,Vangelidis2022}. In addition the Seebeck coefficient is expected to increase in high carrier mobility graphene\cite{Duan2016}. Both effects would give rise to an overall smaller RSM-to-PTE ratio than the simulations presented in \Figrefb{fig:supple_mobility_dependence} for $\tau_p=180\,$ fs and $\tau_p=300\,$ fs. These simulations therefore depict only the upper limits of the expected RSM-to-PTE ratio. The black arrow indicates the mobility dependence of the simulated RSM-to-PTE ratio as a consequence of the changed power dissipation profile in \Figrefa{fig:supple_mobility_dependence}.} The colored dashed vertical lines indicated the high frequency RSM-to-PTE ratio predicted by the CST simulations. A high frequency RSM-to-PTE ratio of roughly 0.2, 0.7 and 1.3 is predicted by the simulations when assuming 60, 180 and 300 fs for $\tau_p$, respectively.
%Building on these computational findings we suggest two main technological outlooks for optimal THz detector performance: (i) Fabrication and design of PTE-based THz detectors should focus on a SAC-type detector layout and CVD-grown graphene as channel material (lower carrier mobility), and (ii) RSM-based THz detectors with AAC-type detector layout should be fabricated with high mobility encapsulated (e.g. in hBN) and exfoliated (monocrystalline) graphene.

\section{Frequency dependence of the extracted RSM and PTE responsivity}

\begin{figure*}[!t]
\centering
\includegraphics[width=6.5in]{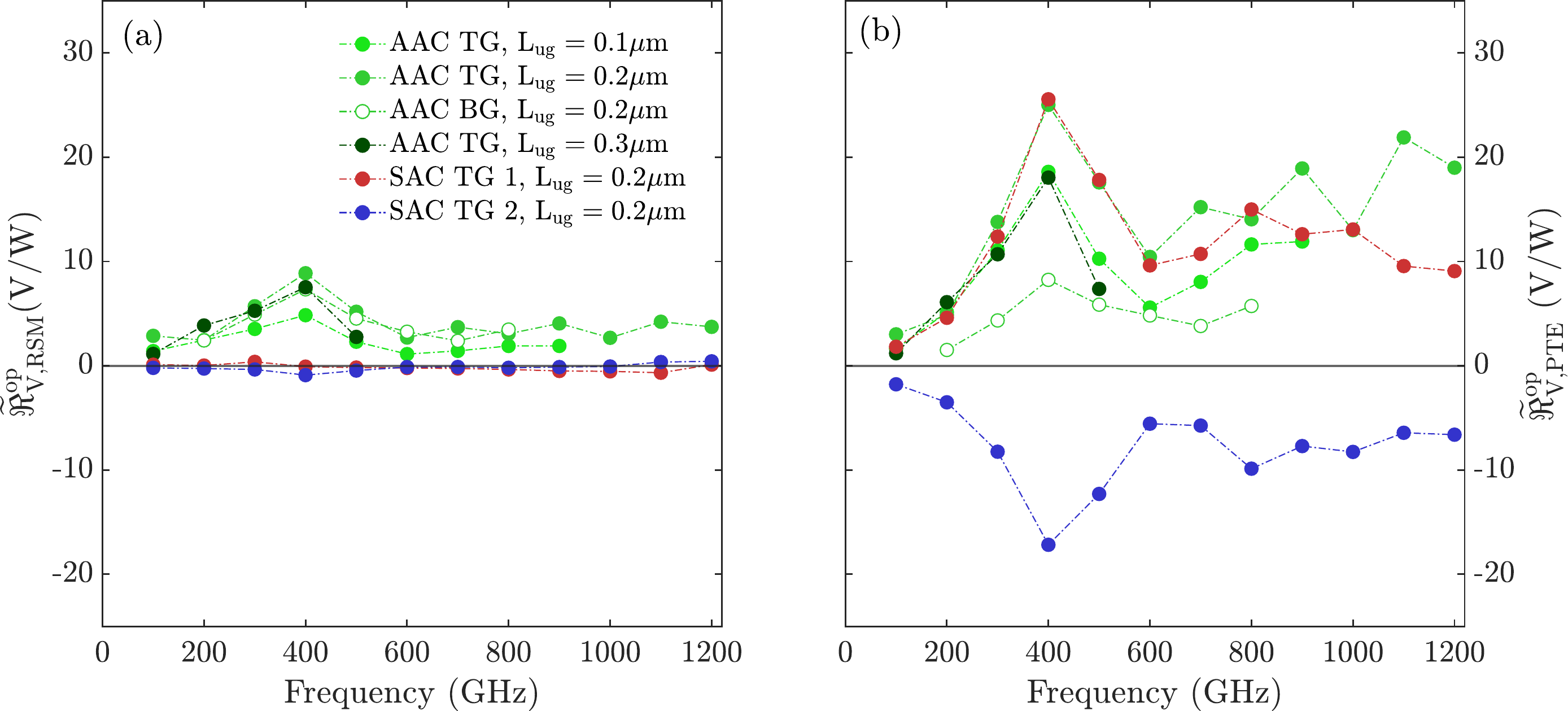}
\caption{\textcolor{black}{(a) Extracted frequency-dependent RSM responsivity ${\widetilde{\Re}}_{V,RSM}^{op}(\omega)$ and (b) PTE responsivity ${\widetilde{\Re}}_{V,RSM}^{op}(\omega)$ for different G-TeraFETs. The extracted responsivites are obtained from the linear fitting routine, which is discussed extensively in the main text. Note that the negative responsivities are due to the opposite sign of the rectified voltage.}}
\label{fig:supple_RSM_PTE_frequency}
\end{figure*}
\textcolor{black}{
In \Figrefa{fig:supple_RSM_PTE_frequency} we plot the extracted RSM (${\widetilde{\Re}}_{V,RSM}^{op}(\omega)$ in (a)) and PTE (${\widetilde{\Re}}_{V,RSM}^{op}(\omega)$ in (b)) for different G-TeraFETs as a function of frequency. Note, that the tilde accent stands here for a special choice of the gate voltage, which was determined from the highest responsivity value at 400~GHz. This gate voltage was then maintained also for all other frequencies depicted (for further details see main text). One finds in \Figrefa{fig:supple_RSM_PTE_frequency}, that for all AAC and SAC photodetectors the efficiency of the PTE recitifcation mechanism dominates that of the RSM at THz frequencies above 100 GHz. At 400 GHz the PTE as well as the RSM responsivity peaks, which is a consequence of the better impedance matching conditions and larger field-enhancement at the respective frequency (see Fig.S6(a) and (b)).}

\section{Extended analysis of the antenna performance}
\begin{figure*}[h!t]
\centering
\includegraphics[width=6.7in]{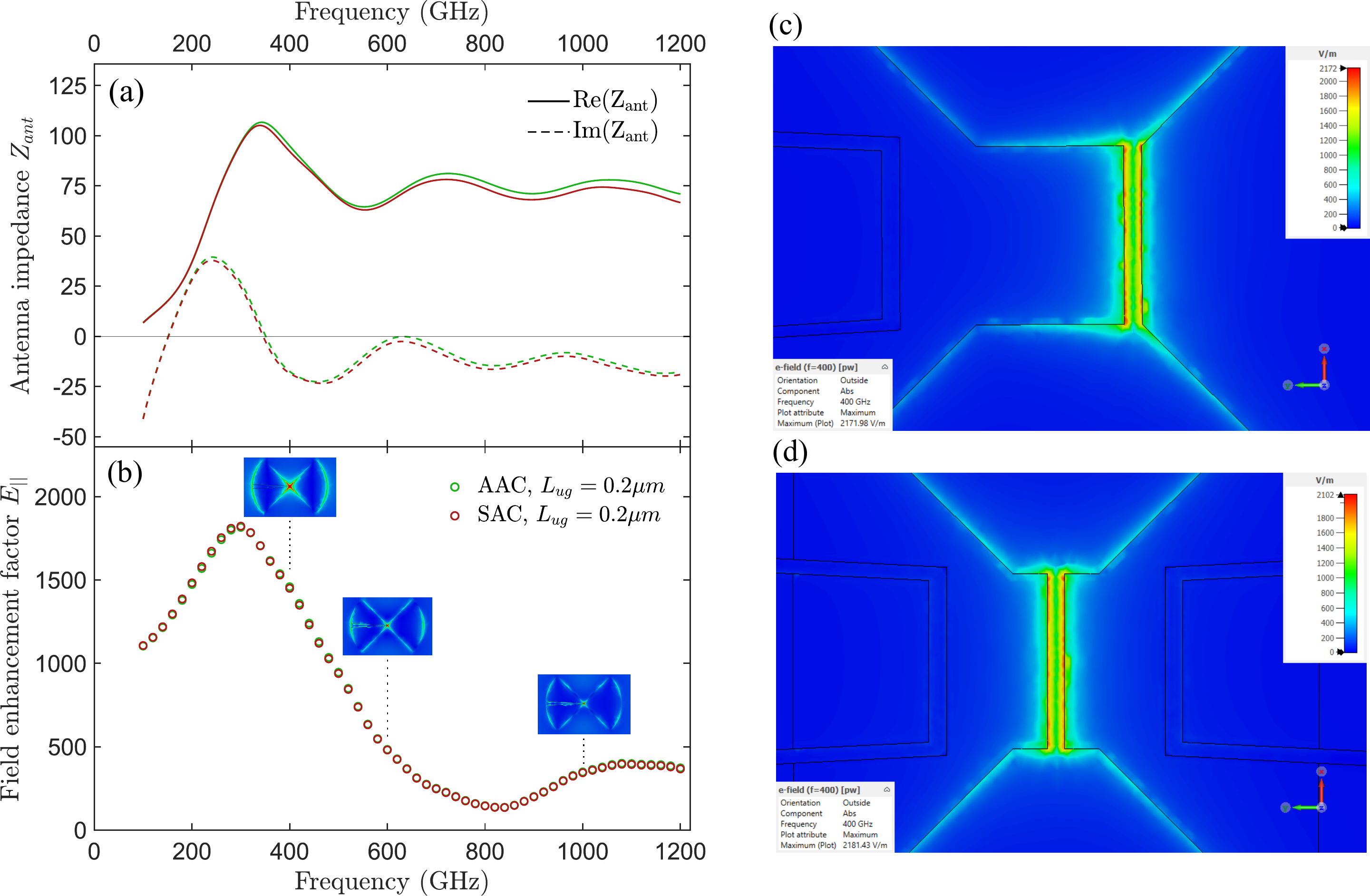}
\caption{\textcolor{black}{(a) Simulated real (solid lines) and imaginary (dashed lines) parts of the antenna impedance for SAC and AAC photodetector layouts with $L_{ug} = 0.2~\mu$m. (b) Extracted field enhancement factor $E_{||}$ obtained in the antenna gap for an incident plane wave excitation polarized alongside $y$ with $E_{ref}=1\,$V/m. The inset pictures in (b) visualize the different field distributions at the respective frequencies. (c)-(d): Two-dimensional plot of the maximum absolute electric field in the antenna gap for AAC (c) and SAC (d) detector layout at 400 GHz. The reference coordinate system is depicted in the bottom right corner.}}
\label{fig:supple_antenna}
\end{figure*}

\textcolor{black}{\Figref{fig:supple_antenna} shows an extensive analysis of the antenna simulations performed for the antenna layouts SAC and AAC, which are studied in this work. For the electromagnetic (EM) wave simulations we used CST Studio Suite. In \Figrefa{fig:supple_antenna} and \Figrefb{fig:supple_antenna} the main figures of merit for the detector antenna performance are presented, which are the complex antenna impedance (presented in (a)) and the field enhancement factor for the in-plane component of the electric field (presented in (b)) in the antenna gap. The latter is calculated from
\begin{align*}
E_{||} = \frac{\max{(E(x,y)|_y)}}{E_{ref}} \;, 
\end{align*}
where $\max{(E(x,y,z)|_y)}$ is the maximum absolute electric field component alongside $y$ (parallel to the field vector of the incident plane wave) and $E_{ref}=1\,$V/m is the reference electric field amplitude of the incident plane wave (oriented alongside $y$). In agreement with our measurement setup the excitation of the antenna is performed from the substrate side (backside illumination). We find that the antenna impedances and the field enhancement factors for the SAC and AAC layouts agree semi-quantitatively, which ensures that the antenna performance is not responsible for the differences in the RSM-to-PTE responsivity ratio roll-offs of both layouts. In \Figrefc{fig:supple_antenna} and \Figrefd{fig:supple_antenna} the maximum absolute field in the antenna gap at 400 GHz is shown. For both layouts the obtained field amplitudes are very similar (see colorbars). The pictures indicate, where heating of the charge carriers in graphene is to be expected from the THz light coupled to the antenna leaves. These EM simulations together with the similar electrostatic properties (see Fig. 1(f) in the main text) and similar channel geometries verify our approach that the detector performances of AAC and SAC can be compared directly with each other.}

\section{Effect of back gate voltage on the AAC detector performance}
\begin{figure*}[h!t]
\centering
\includegraphics[width=6.5in]{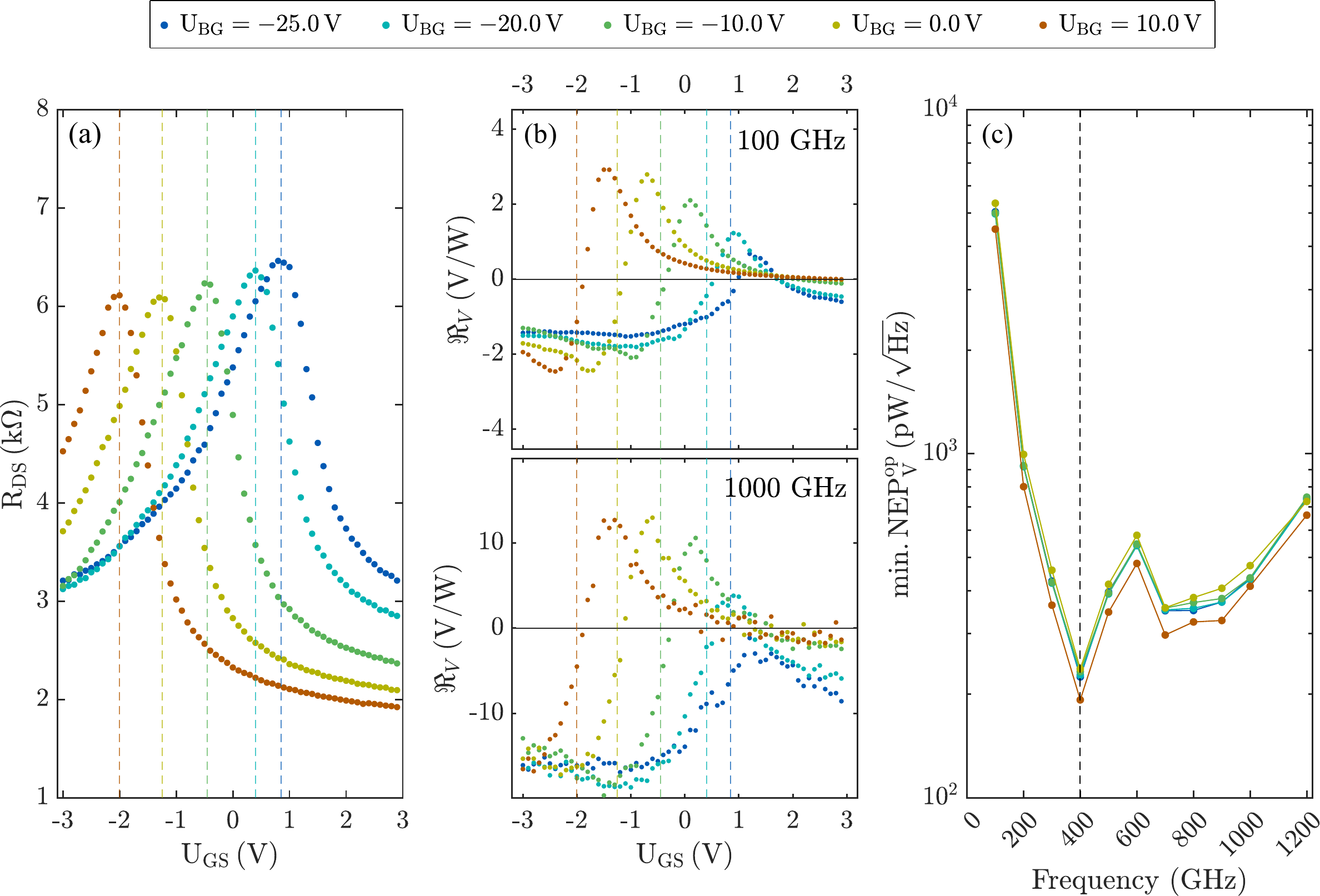}
\caption{\textcolor{black}{(a) Measured drain-to-source resistance $R_{DS}$, (b) voltage responsivity ${{\Re}}_{V}^{op}$ at 100 GHz (top) and 1000 GHz (bottom panel), (c) minimum noise-equivalent power as a function of top gate voltage $U_{GS}$ for detector layout AAC, while fixing the back gate voltage $U_{BG}$ in the range from -25 to 10 V (legend is shown on top of the figure). The vertical dashed lines in (a) and (b) indicate the extracted Dirac voltage $U_{Dirac}$.}}
\label{fig:supple_BG_AAC}
\end{figure*}

\textcolor{black}{
We performed additional DC (see \Figrefa{fig:supple_BG_AAC}) and THz characterization measurements (\Figref{fig:supple_BG_AAC}(b)-(c))) for the AAC detector layout (here $L_{g}=2.5\,\mu$m and $L_{ug}=0.3\,\mu$m) to determine the effect of additional back gate (BG) voltage $U_{BG}$ tuning on the detector performance. In \Figrefa{fig:supple_BG_AAC} the measured drain-to-source resistance $R_{DS}$ as a function of top gate (TG) voltage is shown, while the BG voltage is fixed at different gate voltage ranging from $U_{BG}=10\,$V to $U_{BG}=-25\,$V. It can be observed that for decreasing BG voltage (i) the Dirac voltage shifts from left to right hand side (the channel doping changes from n ($U_{Dirac}<0\,$ V) to p-type ($U_{Dirac}>0\,$ V)), (ii) the overall peak resistance increases and (iii) the Dirac cone gets broadened. The applied BG voltage alters the Seebeck coefficient throughout the full graphene channel region and consequently the PTE signal contribution to the detector signal in the AAC layout is expected to change according to
\begin{equation}
\begin{split}
   & U_{PTE}(U_{GS},U_{BG}) = \\ & \approx \Delta T_C \cdot\left(S_{ug}^{hot}(U_{BG}) - S_g(U_{GS},U_{BG})\right) \,.
    \label{eq:PTEBG}
\end{split}
\end{equation}
In \Figrefb{fig:supple_BG_AAC} the measured voltage responsivity is depicted for the different BG voltages. A clear trend can be observed. For negative BG voltages, e.g. $U_{BG}=-25\,$V, the overall contribution of the PTE to the RSM is increased, which is reflected by an reduced signal magnitude for the electron conduction branch. When comparing the the 100 GHz with the 1000 GHz measurements we observe a similar trend as observed for the devices discussed in the main text, where BG voltage was always set to $U_{BG}=0\,$V. This is, the second phase change on the electron condcution branch with respect to the respective Dirac voltage occurs earlier with increasing frequency (see Fig. 3(d) in the main text), which we could associate to the change in the RSM-to-PTE responsivity ratio resulting from the change in power dissipation along the graphene channel. It is worth mentioning that Eqn.~\ref{eq:PTEBG} predicts a sign change of the PTE when changing from n ($U_{Dirac}<0\,$ V, $S_{ug}^{hot}$<0) to a p-doped ($U_{Dirac}>0\,$ V, $S_{ug}^{hot}$>0) graphene channel, which should here be the case for the measurements at $U_{BG}=-25\,$V and $U_{BG}=-20\,$V. However, contrary to the above equation we find that the PTE contribution to the detector signal is even enhanced for these BG voltages and shows no sign change when switching from n to p-type doping. We attribute this to the discussed contact doping effect (see chapter S4) in close vicinity to the Source metal contact \cite{Chaves2015}. It is known that this effect is penetrating hundreds of nanometers into the graphene channel \cite{Müller2009}, thereby it is to be expected that $S_{ug}^{hot}(U_{BG})$ is partially influenced by the contact doping effect and partially by the back gate voltage. Further investigations are needed to clarify the exact functionality. Finally in \Figrefc{fig:supple_BG_AAC} we plot the extracted minimum Noise-equivalent power $\rm{NEP}^{op}$ as a function of frequency. We observe a similar overall frequency trend as for the samples shown in the main text with the best operation frequency at 400 GHz due to the antenna performance (see chapter S7). We find a minor influence of the BG voltages on the overall detector performance. The best NEP of 193~pW/$\surd{\rm{Hz}}$ is obtained for $U_{BG}=10\,$V.}

\section{Estimation of the prefactors for RSM and PTE}
\textcolor{black}{
We aim to determine the prefactors for the RSM (${U^{el}_{THz}}^2$) and PTE ($\Delta T_{C}$) from the measured free-space THz beam power $P_{THz}^{op}$ (determined in-front of the hyper hemispherical Si lens with a large-area Golay cell) on basis of simple power coupling considerations. The total coupling efficiency of the THz power towards the graphene channel is determined by \cite{Bauer2019,Ferreras2021}
\begin{align}
\eta_{tot} = \frac{P_{Gr}}{P_{THz}^{op}} = \frac{1}{2} \cdot \eta_{op} \cdot \eta_{gauss} \cdot \eta_{ant} (\omega) \cdot \eta_m (\omega) \,,
\label{eq:power_coupling}
\end{align}
where $P_{Gr}$ is the fraction of the total free-space beam power $P_{THz}^{op}$ which is collected by the antenna and transmitted to the graphene channel. The factor of 1/2 incorporates residual scattering of incident power in the antenna element \cite{Andersen2005}. $\eta_{gauss}$ represents the \textit{gaussicity} \cite{Filipovic1993} - a quantity typically employed for substrate lens antennas and is assumed to be 0.9 \cite{Filipovic1993,Bauer2019,Ferreras2021}. $\eta_{ant} (\omega)$ is the frequency-dependent antenna efficiency, which is simulated by CST Studio Suite. $\eta_{op} \sim 0.46$ incorporates all optical path losses consisting of reflections of the incident THz light at the air-to-silicon-lens interface of the detector (0.7)\cite{Bauer2019} and Drude absorption in the lightly p-doped substrate (0.66, determined experimentally, see chapter S2). Finally, $\eta_m (\omega)$ can be obtained from \cite{Boppel2012,Bauer2019}}
\textcolor{black}{
\begin{align}
\begin{split}
\eta_m (\omega) &= \frac{4 \cdot \rm{Re}[Z_{ant}(\omega)] \cdot  \rm{Re}[Z_{Gr}]}{|Z_{ant}(\omega)+Z_{Gr}|^2}  
\end{split}
\end{align}
and accounts for power losses due to the impedance mismatch between the antenna element ($Z_{ant}(\omega)$, shown in \Figrefa{fig:supple_antenna}) and the graphene channel $Z_{Gr}$. For the calculation we assume the latter to be constant in frequency (determined from the measured DC resistance, $Z_{Gr} \approx 3\, \rm{k \Omega}$). 
With Eqn.~\ref{eq:power_coupling} $U^{el}_{THz}$ can determined from the power of the incident THz radiation via
\begin{align}
U^{el}_{THz} = \sqrt{P_{Gr} \cdot Z_{Gr}}\,.
\label{eq:UTHz}
\end{align}
In order to estimate the carrier temperature increment $\Delta T_{C} =T_C - T_L$ of the carrier temperature $T_C$ with regards to the lattice temperature $T_L$ in the antenna gap we use the steady-state carrier-heating equation\cite{Massicotte2021}
\begin{align}
\begin{split}
\Delta T_C = T_C - T_L = \frac{P_{A,in} \tau_{c}}{C_e} \,, 
\end{split}
\label{eq:temp_increament}
\end{align}
where $P_{A,in}={P_{Gr}}/A_{gap}$ is the absorbed power density in the antenna gap ($A_{gap}$=$L_{ug} \cdot W$, $L_{ug}= 0.2 \, \mu$m, $W = 2 \, \mu$m), $\tau_{c}$ is the cooling time, which can take place on a time scale $\tau_c\,{\lesssim}\,1$~ps \cite{Gierz13,Frenzel14,Aamir21} and is assumed to be 1 ps. $C_e$ is the electronic heat capacity. In the doped regime $C_e$ it is given by\cite{Massicotte2021}
\begin{align}
\begin{split}
C_e = \frac{2\pi E_F}{3 (\hbar v_F)^2} k_B^2 T_C = \gamma_{doped} \cdot T_C \,.
\end{split}
\label{eq:heat_capacity}
\end{align}
}
\begin{figure*}[!t]
\centering
\includegraphics[width=5in]{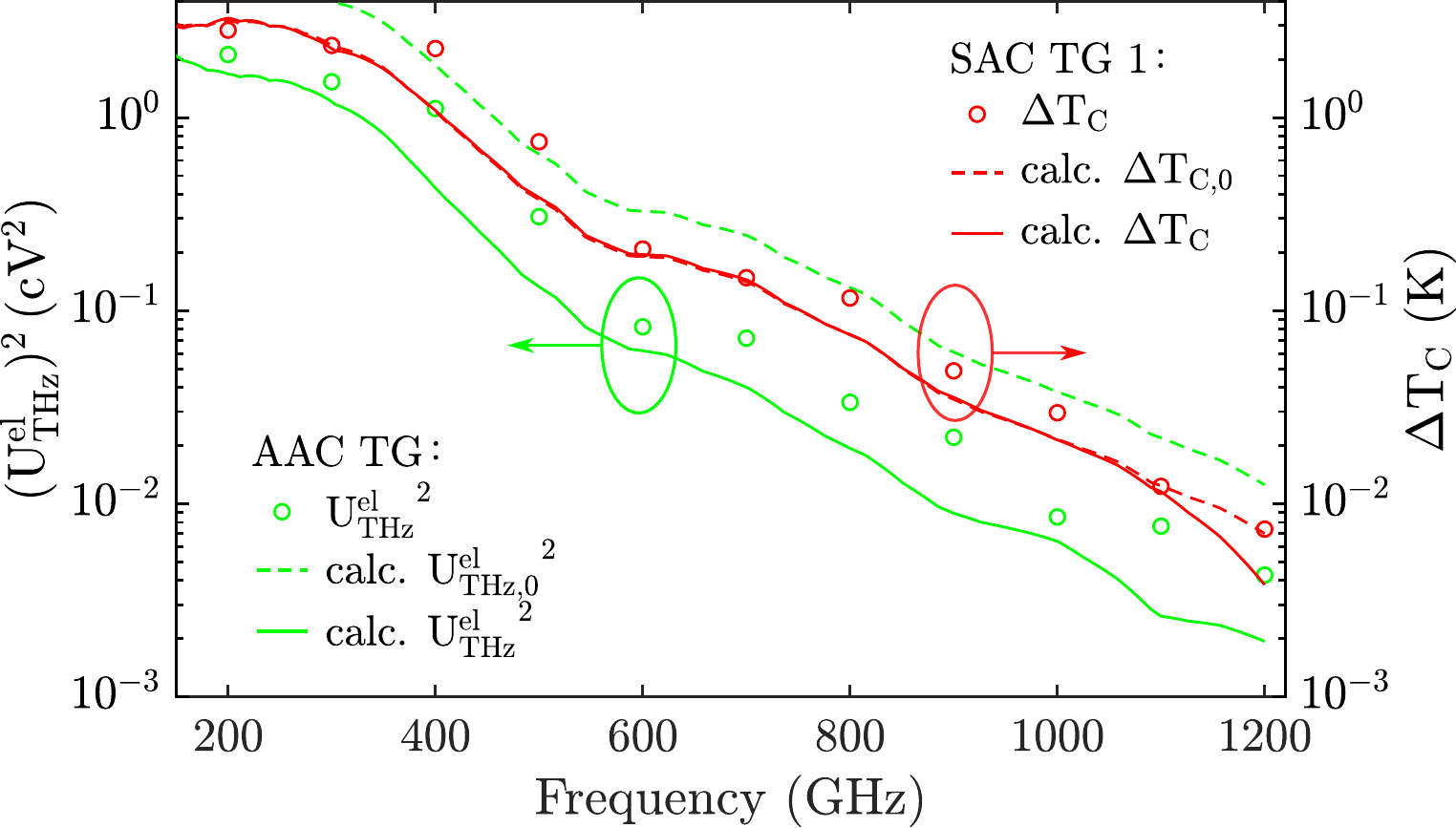}
\caption{\textcolor{black}{Extracted prefactors for the (${U^{el}_{THz}}^2$, in green, left axis) and PTE ($\Delta T_{C}$, in red, right axis) signal contributions in the G-TeraFETs presented in the main text with $L_{ug}= 0.2 \, \mu$m in comparison with estimations, which have been determined from the measured free-space THz beam power $P_{THz}^{op}$.}}
\label{fig:supple_prefactors}
\end{figure*}
\textcolor{black}{
Here $E_F$ is the Fermi level (we assume a constant value of $E_F=0.1\,$eV), $v_F = 1.1\cdot 10^6$ is the Fermi velocity and $\hbar$ is the reduced Planck constant. Combining equation Eqn.~\ref{eq:temp_increament} and Eqn.~\ref{eq:heat_capacity} we obtain
\begin{align}
\begin{split}
\Delta T_C = \frac{1}{2}\cdot \left(\frac{\sqrt{\gamma_{doped}\cdot T_L^2+4 P_{A,in} \tau_c}}{\sqrt{\gamma_{doped}}} - T_L\right) \,. 
\end{split}
\label{eq:dT}
\end{align}
In \Figref{fig:supple_prefactors} the final extracted prefactors for the AAC (green, left axis) and SAC (red, right axis) photodetector layout are shown in comparison with the respective calculations using Eqn.~\ref{eq:UTHz} for AAC (calc. ${U^{el}_{THz,0}}$, dashed green line) and Eqn.~\ref{eq:dT} for SAC (calc. $\Delta T_{C,0}$, red dashed line). In agreement with the extracted prefactors obtained from the THz characterization measurement and to account for the frequency-dependent change in power consumption of the RSM and PTE process in each detector layout we then determine the final prefactors from
$({U^{el}_{THz}})^2 = X_{RSM} \cdot ({U^{el}_{THz,0}})^2$ (solid green line) and $\Delta T_{C} = X_{PTE} \cdot \Delta T_{C,0}$ (solid red line). We find a semi-quantitative agreement between the calculated and extracted prefactors, indicating that the fitting approach used in this work provides reasonable physical orders of magnitude for the prefactors for both rectification processes. Note that deviations between calculations and extracted prefactors are to be expected, as we are not taking the frequency-dependent graphene channel impedance into account and assume it to be constant in frequency and real-valued.}

\end{document}